\documentclass[epj, nopacs]{svjour}
\usepackage{epsfig}
\newcommand{\be}{\begin{equation}}
\newcommand{\ee}{\end{equation}}
\newcommand{\bea}{\begin{eqnarray}}
\newcommand{\eea}{\end{eqnarray}}
\begin{document}

\title{ $\alpha_s$ from $\tau$ decays:  contour-improved versus fixed-order summation in a new QCD perturbation expansion}

\author{ Irinel Caprini\inst{1} and Jan Fischer\inst{2}}
\institute{National Institute of Physics and Nuclear Engineering, \\POB MG 6,
Bucharest, R-077125 Romania  \and
Institute of Physics, Academy of Sciences of the Czech Republic, \\
CZ-182 21  Prague 8, Czech Republic}

\date{Received: date / Revised version: date}
\abstract{We consider the determination of  $\alpha_s$ from $\tau$ hadronic decays, by investigating the contour-improved (CI) and the fixed-order (FO) renormalization group summations  in the frame of a new perturbation expansion of QCD, which incorporates in a systematic way the available information about the  divergent character of the series. The new expansion  functions, which replace the powers of the coupling,  are defined by the analytic continuation in the Borel complex plane, achieved through an optimal conformal mapping. Using a physical model recently discussed by Beneke and Jamin, we show that the new CIPT approaches the true results  with great precision when the perturbative order is increased, while the new FOPT gives a less accurate description in the regions where the imaginary logarithms present in the expansion of the running coupling   are large. With the new expansions, the discrepancy of  0.024 in $\alpha_s(m_\tau^2)$  between the standard CI and FO summations is  reduced to only 0.009.  From the new CIPT  we predict $\alpha_s(m_\tau^2)=  0.320\,\, ^{+0.011}_{-0.009}$, which practically coincides with the result of the standard FOPT, but has a more solid theoretical basis.} 

\authorrunning{I.Caprini and J. Fischer}
\titlerunning{$\alpha_s$ from $\tau$ decays}
 \maketitle

\section{Introduction} \label{intro}
The precise determination of $\alpha_s$ from $\tau$ hadronic decays is one of the most important results in perturbative QCD \cite{PDG2008}. The subject has been treated by many authors (see \cite{Bra88}-\cite{Maltman08} and references therein).  Among the most important improvements we mention the so-called "contour-improved renormalization-group summation" \cite{Pivo}, \cite{DiPi}, which avoids the large logarithms of the usual "fixed-order" expansion along the integration contour relevant for $\alpha_s$ extraction. The  higher orders of perturbation theory, expressed by the so-called renormalons, were also investigated, especially as concerns their effect on the precision of the theoretical determinations \cite{BBB}, \cite{Neub}. 

The problem was revisited  recently, after the  calculation of the Adler function up to fourth order  \cite{BCK08}, the same at which 
  the $\beta$ function describing the running of the coupling is known \cite{LaRi}, \cite{Czakon}. The determination of $\alpha_s$ from the ALEPH spectral function data was reconsidered in \cite{Davier2008}, where the prediction based on CIPT  $\alpha_s(m_\tau^2)=0.344\pm 0.009$ was made. On the other hand, an updated version of   FOPT \cite{BeJa}
 led to the prediction  $\alpha_s(m_\tau^2)=0.320^{+0.012}_{-0.007}$.   As remarked in \cite{BeJa},  the discrepancy of 0.024  between  CIPT and FOPT appears to be the largest systematic theoretical uncertainty in the $\alpha_s$ determination, and it does not go away by adding the presently known higher-order terms.

 The two summation methods were analyzed in detail in \cite{BeJa} by means of  a "physical" model for the Borel transform of the Adler function. The conclusion of this study was that, somewhat surprisingly,   FOPT is preferable,  CIPT failing to approach the true result, although it is a priori more consistent.

 The purpose of the present work is to understand better the relation between 
the CI and FO summation methods and their consequences on the extraction of $\alpha_s$. To this end we investigate these two  methods  using a new perturbation expansion advocated by us some time ago \cite{CaFi}. The new expansion separates the intrinsic ambiguity  of the  perturbation theory due to the infrared regions of the Feynman diagrams, from the divergent character of the series. This is achieved if the usual perturbative expansion is replaced by a series with better convergence properties (each new term improving the accuracy of the
approximation instead of spoiling it).  As shown in \cite{CaFi}-\cite{CaFi2}, the new expansion is defined by  using the  analytic continuation in the Borel complex  plane 
 by an optimal conformal mapping.  The properties of the novel expansion functions  were analyzed in detail in \cite{CaFi1},\cite{CaFi2} and some applications were discussed in \cite{CaFi}, \cite{ChFi}.

The paper is organized as follows: in Section \ref{sec:CIFO} we briefly review the CI and FO summations of the Adler function relevant for  $\tau$ decays. In Section \ref{sec:borel} we discuss the Borel transform and in Section \ref{sec:new} we present  the CI and FO versions of the new perturbation theory for the Adler function. In Section \ref{sec:alphas} we apply  these two methods to calculate  $\alpha_s(m_\tau^2)$, and show that the difference between the two predictions is significantly smaller than in the standard case. In order to understand these results, in Section \ref{sec:BJ} we investigate   the new expansions using as an example the physical model proposed in \cite{BeJa}.  More comments on the method of conformal mapping and its relevance for the physical case are made in Section \ref{sec:disc}. 
We summarize our results in Section \ref{sec:conc}, where we emphasize that 
the  CI summation combined with the new perturbation expansion leads to a precise  determination of $\alpha_s(m_\tau^2)$.

\section{Standard CIPT and FOPT \label{sec:CIFO}}
The relevant quantity for the extraction of $\alpha_s(m_\tau^2)$ is the integral
\be\label{delta0} \delta^{(0)} =  \frac{1}{2\pi i}\, \oint\limits_{|s|=s_0}\, \frac{d s}{s}\, \omega(s)\,\hat D(s)\,,
\ee
where $s_0=m_\tau^2$,  $\omega(s)=1-2s/s_0 + 2(s/s_0)^3-(s/s_0)^4$  and $\hat D(s)=D(s)-1$ is the reduced Adler function  in
massless QCD. It is written  formally as  the renormalization-group improved series\footnote{The normalization  of $D$ is that adopted in \cite{BeJa}, where  $K_n$ are denoted as $c_{n,1}$.  For simplicity,   in (\ref{DCI})  the scale was set to $\xi=1$. The general case will be discussed at the end of Section \ref{sec:new}.}
  \begin{equation}\label{DCI} 
\hat D(s) =   \sum_{n=1}^{\infty} K_{n}\,  (a_s(s))^n\,, \end{equation} 
where $a_s(s) \equiv \alpha_s(s)/\pi$.
   In the $\overline{MS}$ scheme, for $n_f=3$,  the coefficients $K_n$  calculated up to now have the values:
\be\label{Ki}
K_1=1,~~ K_2= 1.6398,~~  K_3=6.3712,~~ K_4=49.076.
\ee
Several methods  were proposed for estimating  the higher-order perturbative coefficients from the low-order ones  \cite{KaSt}, \cite{BCK03}. In particular, in \cite{BeJa} the authors adopt the value
\be\label{K5}
K_5=283.
\ee 
We mention that at large $n$ the coefficients display a factorial increase, $K_n\sim n!$,  so that the series in  (\ref{DCI}) is divergent.  In writing  (\ref{DCI})  we follow the convention often adopted in physical papers, writing the sign of equality even if the series on the right hand side is divergent and the equality is impossible. According to Dyson's proposal  \cite{Dyson} from 1952, the series is then regarded as asymptotic to $\hat D(s)$ standing on the left hand side. If the series is convergent, the relation (\ref{DCI}) is understood as equality. Analogous series in the text below (see (\ref{DnewCI0}), (\ref{DnewCI}), (\ref{DnewFO0}), (\ref{DnewFO}), etc.) are understood in a similar sense. 

The contour-improved (CI) summation amounts to introducing (\ref{DCI}) in (\ref{delta0}), and performing the integral with $a_s(s)$ calculated locally from the solution of the renormalization group equation. The solution is known at present up to four loops \cite{LaRi}, the first coefficients  $\beta_j$  of the $\beta$ function,  calculated in the  $\overline{MS}$ scheme and $n_f=3$  being
\be\label{betai}
\beta_0=9/4,~~\beta_1=4,~~\beta_2=10.0599,~~\beta_3=47.228.
\ee 

The Taylor expansion of $a_s(s)$ in terms of a reference point $s_1$ reads
\be\label{astaylor}
a_s(s)=\sum\limits_{j\ge 1}\, \xi_j\, (a_s(s_1))^j,
\ee
where the coefficients  $\xi_j$  depend on $\eta_1=\ln (s/s_1)$ and on the coefficients $\beta_j$. 
By inserting this expansion with $s_1=s_0$ in (\ref{DCI}) and  rearranging the expansion  in powers of $a_s(s_0)$,  one obtains the FO summation, which can be written as
\be\label{DFO}
\hat D(s) =   \sum_{n=1}^{\infty} [K_{n}+\kappa_n(s)]\,  (a_s(s_0))^n\,,
\ee
where $\kappa_n(s)$  are polynomials  of $\eta_0=\ln (s/s_0)$, and depend on the coefficients $\beta_j$ and  $K_j$ for $j\le n$.

\section{Borel transform}\label{sec:borel}
The series (\ref{DCI}) can be formally written as  the  Borel-Laplace transform
\begin{equation}\label{Laplace}
  \hat D(s)=\frac{1}{\beta_0}\,\int\limits_0^\infty\!{\rm e}^{-u/(\beta_0 a_s(s))} \, B(u)\,{\rm d}u\,,\end{equation}
where 
 $B(u)$ is the Borel transform of the Adler function, defined by the power series
\begin{equation}\label{B}  B(u )=\sum\limits_{n=0}^\infty b_n u^n\,, 
\end{equation} 
 with  $b_n$ related to the  original
perturbative  coefficients appearing in (\ref{DCI}) by 
\begin{equation}\label{bn}
b_n=\frac{K_{n+1}}{\beta_0^n \,n!}\,,\quad n\ge 0. \end{equation}  
 According to present knowledge, the function $B(u)$ has   branch
point  singularities  in the $u$-plane, along the negative
axis - the ultraviolet (UV) renormalons - and the positive axis - the  infrared
 (IR) renormalons \cite{Beneke}. Specifically, the branch cuts are situated  along the rays $ u
\leq -1$ and $u \geq 2$.  The nature of the first
branch points  was established in \cite{Muel} and in \cite{BBK} (see also \cite{BeJa}). Thus, near the first branch points, {\em i.e.} for $u\sim -1$ and $u\sim 2$, respectively,  $B(u)$  behaves as
\be\label{branch}
B(u)\sim \frac{r_1}{(1+u)^{\gamma_1}},\quad B(u)\sim \frac{r_2}{(1-u/2)^{\gamma_2}},
\ee
where the residues $r_1$ and $r_2$ are not known, but the exponents $\gamma_1$ and $\gamma_2$ are known \cite{Muel}, \cite{BBK}.

Due to the singularities along the positive axis, the integral (\ref{Laplace}) does not exist. The ambiguity in the choice of prescription is often used as a measure of the uncertainty of the calculations in perturbative QCD.
 It is convenient to define the integral by the Principal Value (PV) prescription:
\be\label{Dpv}
\hat D(s)\equiv\frac{1}{\beta_0}\,{\rm PV}\int\limits_0^\infty\!{\rm e}^{-u/(\beta_0 a_s(s))} \, B(u)\,{\rm d}u\,,
\ee
where
\bea\label{Pv}
{\rm PV} \int\limits_0^\infty f(u){\rm d}u  \equiv  \frac{1}{2}\int\limits_0^\infty \left[ f(u+i \epsilon){\rm d}u + f(u-i \epsilon){\rm d}u\right],~~ \epsilon \to 0. \nonumber
\eea
As discussed in \cite{CaNe}, \cite{CaFi3}, the PV  prescription  is the best choice  if one wants to preserve as much as possible  the analyticity properties of the correlators in the  $s$-plane, which are connected with causality and unitarity.

\section{New CIPT and FOPT \label{sec:new}}
In order to define a new perturbative expansion of the Adler function we shall apply the method of conformal mapping \cite{CiFi}.
This method  is not applicable to the series (\ref{DCI}), because 
${\hat D}(s)$ (regarded as a function of $a_{s}(s)$) is singular at the point of 
expansion $a_{s}(s)=0$. The method can, on the other hand, be applied to 
(\ref{B}), because $B(u)$ is holomorphic at $u=0$. 

We note that  the expansion (\ref{B})  converges only in the disk $|u|<1$. A series  with a larger domain of convergence can be obtained by expanding $B(u)$ in powers of a new variable.   As shown in \cite{CiFi}, the optimal variable coincides with the function that performs the conformal mapping of the whole analyticity domain of the expanded function onto a disk in the new complex plane\footnote{For QCD, the use of a conformal mapping in the Borel plane  was suggested in \cite{Muel1} and was applied in a more limited context in \cite{Alta}. Applications of the method  were considered also in \cite{Jent}, \cite{CvLe}.}. 

To find the explicit form of the optimal conformal mapping, one
should know the location of all the singularities of the Borel transform
$B(u)$ in the complex Borel plane. Unfortunately, present evidence of 
these singularities is  scarce:  the known
singularities (IR and UV renormalons and instanton-antiinstanton
pairs) are produced only by a subclass of Feynman diagrams, while
the effect of all the  diagrams on the nature and location 
the singularities is not known. In the lack of rigorous
results, additional assumptions are made or special models are  built. As for additional assumptions, we point out that universally
adopted has been to assume that $B(u)$ has only the above-mentioned singularities on the real axis with a
gap, being holomorphic elsewhere.  

Under this assumption, the optimal variable  defined in \cite{CiFi}  reads \cite{CaFi}:
\be\label{w}
w(u)=\frac{\sqrt{1+u}-\sqrt{1-u/2}}{\sqrt{1+u}+\sqrt{1-u/2}}.
\ee
This function maps the $u$-plane cut for $u\ge 2$ and $u\le -1$ onto the unit disk $|w|<1$  in the complex plane $w=w(u)$, such that $w(0)=0$,  $w(2)=1$ and  $w(-1)=-1$. It is useful to give also the inverse $u=u(w)$  of (\ref{w}): 
\be\label{uw}
u(w)=\frac{8 w}{3 - 2 w + 3 w^2}. 
\ee
According to general arguments \cite{CiFi},   the expansion
 \be\label{Bw0}
B(u)=\sum_{n\ge 0} d_n \,w^n
\ee
converges in the whole disk $|w|<1$.  Moreover, as shown in \cite{CiFi}, the expansion (\ref{Bw0}) has the best asymptotic rate of convergence compared to all the expansions of the function $B(u)$ in powers of other variables. 

The series (\ref{Bw0}) can be used to define an alternative  expansion of $\hat D(s)$. This is obtained formally by inserting (\ref{Bw0}) into  (\ref{Dpv}) and interchanging the order of summation and integration. Thus, we adopt the modified CIPT expansion  defined as \cite{CaFi}-\cite{CaFi2}
\be\label{DnewCI0}
\hat D(s)=\sum\limits_{n\ge 0} d_n W_n(s),
\ee
where
\be\label{Wn0}
W_n(s)=\frac{1}{\beta_0}{\rm PV} \int\limits_0^\infty\!{\rm e}^{-u/(\beta_0 a_s(s))} \, w^n\,{\rm d}u\,.
\ee
We empahsize that the expansion (\ref{Bw0}) exploits only  the location of  the  singularities in the Borel plane.  However, in our case some information exists also about the nature of the singularities.  We note that Eq. (\ref{branch}) expressed in the variable $w=w(u)$  implies
\be\label{branchw}
B(u)\sim \frac{(8/3)^{\gamma_1}r_1 }{(1+w)^{2\gamma_1}},\quad
\quad B(u)\sim \frac{(4/3)^{\gamma_2} r_2}{(1-w)^{2\gamma_2}},
\ee
for the behaviour near the points $w=-1$ and  $w=1$, respectively.
The  expansion (\ref{Bw0}) is expected to describe these singularities  if a large number of terms is used.  However, since the nature
 of the singularities is known, it is convenient to  incorporate it  explicitly.  This is achieved, for instance, by expanding the product $(1+w)^{2\gamma_1}(1-w)^{2\gamma_2}\, B(u)$ in powers of the variable $w$:
 \be\label{Bwgamma}
(1+w)^{2\gamma_1}(1-w)^{2\gamma_2} B(u)= \sum\limits_{n\ge 0} c_n \,w^n.
\ee
The expansion (\ref{Bwgamma})  converges in the whole disk  $|w|<1$, {\em i.e.} in the whole complex $u$-plane, up to the cuts along the real axis. Moreover, since the singular behaviour of $B(u)$ at the first branch points is compensated by the first factors in  (\ref{Bwgamma}), the series is expected to converge faster than (\ref{Bw0}). Also, the behaviour near the first singularities holds even for truncated expansions, which are used in practice.

It is important to stress that, while the expansion (\ref{Bw0}) is unique, the explicit inclusion of the first singularities of $B(u)$  contains some arbitrariness. The description of the singularities by multiplicative factors is a possibility, but is  not a priori necessary. Moreover, the factors are not unique.  For instance, the information on the nature of the singularities can be exploited by  factors in the $u$ variable\footnote{More precisely, in \cite{Soper} the product  $(1+u)^{\gamma_1}(1-u/2)^{\gamma_2}B(u)$ was expanded  in powers of $u$, while in \cite{ChFi} the same product was expanded in powers of $w$.}.  An  advantage of the choice  (\ref{Bwgamma}) is that the multiplicative factors  remain finite  at large $u$.  We will make more comments on this in  Section \ref{sec:disc}.

The expansion (\ref{Bwgamma}) suggests  the definition of the new CIPT
\be\label{DnewCI}
\hat D(s)=\sum\limits_{n\ge 0} c_n {\cal W}_n(s), 
\ee
where the expansion functions are defined as
\be\label{Wn}
{\cal W}_n(s)=\frac{1}{\beta_0}{\rm PV} \int\limits_0^\infty\!{\rm e}^{-u/(\beta_0 a_s(s))} \,  \frac{w^n}{(1+w)^{2\gamma_1}(1-w)^{2\gamma_2}}\,{\rm d}u\,,
\ee
with $w=w(u)$ defined in (\ref{w}).

We emphasize that the expansions (\ref{DnewCI0}) and  (\ref{DnewCI}) reproduce the coefficients $K_n$ of the usual expansion (\ref{DCI}), when the  functions (\ref{Wn0}) and (\ref{Wn}) are expanded  in powers of the coupling.
In fact, as shown in \cite{CaFi2}, the new expansion functions  are formally represented by divergent series  in powers of the coupling, much like the expanded correlator itself. 

 To obtain the FO version of the new expansions, we start from (\ref{DFO}) and 
define the corresponding Borel transform    
\begin{equation}\label{BFO} 
 \tilde B(u,s )=\sum\limits_{n=0}^\infty \tilde b_n(s) u^n\,, 
\end{equation} 
 where
\begin{equation}\label{bnFO}
\tilde b_n(s)=\frac{K_{n+1}+\kappa_{n+1}(s)}{\beta_0^n\, n!}\,,\quad n\ge 0. \end{equation}  
Then the Adler function admits the formal representation
\begin{equation}\label{LaplaceFO}
  \hat D(s)=\frac{1}{\beta_0}\,\int\limits_0^\infty\!{\rm e}^{-u/(\beta_0 a_s(s_0))} \, \tilde B(u,s)\,{\rm d}u\,.\end{equation}
By comparing Eqs.(\ref{BFO}), (\ref{bnFO}) with (\ref{B}), (\ref{bn}), we can write 
\be\label{BB1}
\tilde B(u,s)=B(u) + B_1(u,s),
\ee
where  $B_1(u,s)$ is generated by  the second term  in the coefficients (\ref{bnFO}).
It follows that the singularities of  $B(u)$  are present also in the function  $\tilde B(u,s)$, which may have in addition singularities from the second term in (\ref{BB1}).  In what follows we shall exploit the singularities of  
$B(u)$, which are known, by expanding  $\tilde B(u, s)$ in powers of the variable $w$ defined above:
\be\label{BwFO}
 \tilde B(u, s)= \sum\limits_{n\ge 0} \tilde d_n(s) \,w^n.
\ee
This leads us to the definition of a modified  FOPT, analogous to the CIPT expansion (\ref{DnewCI0}):
\be\label{DnewFO0}
\hat D(s)=\sum\limits_{n\ge 0} \tilde d_n(s)\, \tilde{W}_n,
\ee
in terms of the functions
\be\label{WnFOw0}
\tilde{W}_n=\frac{1}{\beta_0}{\rm PV} \int\limits_0^\infty\!{\rm e}^{-u/(\beta_0 a_s(s_0))} \, w^n \,{\rm d}u\,.
\ee
 As we discussed above,  it is convenient to impose explicitly  the behaviour (\ref{branch}), which is done by expanding
\be\label{BwgammaFO}
(1+w)^{2\gamma_1}(1-w)^{2\gamma_2} \tilde B(u, s)= \sum\limits_{n\ge 0} \tilde c_n(s) \,w^n.
\ee
Then, using (\ref{LaplaceFO}) we  define the new  FO expansion:
\be\label{DnewFO}
\hat D(s)=\sum\limits_{n\ge 0} \tilde c_n(s)\, \tilde{\cal W}_n,
\ee
in terms of the functions
\be\label{WnFO}
\tilde{\cal W}_n=\frac{1}{\beta_0}{\rm PV} \int\limits_0^\infty\!{\rm e}^{-u/(\beta_0 a_s(s_0))} \,  \frac{w^n}{(1+w)^{2\gamma_1}(1-w)^{2\gamma_2}}\,{\rm d}u\,.
\ee

In the next Section we shall determine $\alpha_s$ using  the new CI and FO  perturbation expansions (\ref{DnewCI}), (\ref{Wn}) and (\ref{DnewFO}), (\ref{WnFO}), which include in an explicit way the nature of the first singularities of the Borel transform. More comments on them and the  general expansions  (\ref{DnewCI0}), (\ref{Wn0}) and (\ref{DnewFO0}), (\ref{WnFOw0}), which replace the standard CIPT and FOPT, respectively,  will be made  in Section \ref{sec:disc}.

 By inserting (\ref{DnewCI}) in the integral (\ref{delta0}), with $a_s(s)$ calculated from RGE applied  locally, we obtain a new CI perturbation expansion for $\delta^{(0)}$. Likewise, by using in the integral (\ref{delta0})  the expansion 
(\ref{DnewFO}),  we obtain the new FO perturbation expansion of $\delta^{(0)}$.

We end this Section with a comment about the renormalization  scale. The starting point of our derivation was the renormalization group improved expansion (\ref{DCI}), which  corresponds to the choice of the scale $\xi=1$ (with the notation used in \cite{Davier2008}). The general case 
  \begin{equation}\label{DCIscale} 
\hat D(s) =   \sum_{n=1}^{\infty} \hat K_{n}(\xi)\,  (a_s(\xi s))^n\,, \end{equation} 
with the coefficients  $\hat K_{n}(\xi)$ given in \cite{Davier2006}, is easily obtained by reordering (\ref{DCI}) as a series in powers of  $a_s(\xi s)$. Similarly, the more general version of the FOPT
\begin{equation}\label{DFOscale} 
\hat D(s) =  \sum_{n=1}^{\infty} [\hat K_{n}(\xi)+ \hat \kappa(s, \xi)]\,  (a_s(\xi s_0))^n\,, \end{equation} 
is obtained by reordering (\ref{DCIscale}) in powers of  $a_s(\xi s_0)$.

Starting from (\ref{DCIscale}) and  (\ref{DFOscale}), the new CI and FO expansions for an arbitrary scale can be obtained in a straightforward way. Both the expansion coefficients and the  expansion functions appearing in the
 generalizations of (\ref{DnewCI}) and 
(\ref{DnewFO}) will depend on the scale $\xi$.
 
\section{Determination of $\alpha_s(m_\tau^2)$}\label{sec:alphas}

Our objective is to compare the results of the standard CIPT and FOPT from \cite{Davier2008}, \cite{BeJa}, with the predictions of the new perturbation expansion presented in Section \ref{sec:new}. To facilitate the comparison, we shall adopt the phenomenological value quoted in \cite{BeJa}:
\be\label{phen}
\delta^{(0)}_{\rm phen}=0.2042 \pm 0.0050.
\ee
The determination of $\alpha_s(m_\tau^2)$ then amounts to the calculation of $\delta^{(0)}$ defined in (\ref{delta0}) using a specific expansion of $\hat D(s)$ for $s_0=m_\tau^2$, and  solving  the equation
$\delta^{(0)}=\delta^{(0)}_{\rm phen}$ with respect to the coupling.

We use the known coefficients $K_n$ from (\ref{Ki}). For $K_5$  we adopt the central value (\ref{K5}) with an error of $\pm $ 50\%. Moreover, we follow the analysis made in  \cite{BeJa}, based on earlier works \cite{Muel}, \cite{BBK}, \cite{Beneke}, which leads to:
\be\label{gammai}
\gamma_1= 1.21,\quad \quad \gamma_2=2.58.
\ee
The Taylor coefficients $b_n$ of the Borel transform, defined in  (\ref{bn}), are
\bea\label{bnnum}
&&b_0=1, \quad  b_1= 0.7288,  \quad  b_2=0.6292 \nonumber\\
&&b_3=0.7181, \quad b_4=0.4601 .
\eea
Then the coefficients $c_n$ appearing in the new expansion (\ref{Bwgamma}), truncated at $n\le 4$ read:
\bea\label{cnnum}
&&c_0=1,\quad c_1=-0.7973, \quad c_2=0.4095,\nonumber\\
&& c_3=8.6647,\quad  c_4=2.2416.
\eea
The  new CIPT  is given by the expansion (\ref{DnewCI}) truncated after $N=5$ terms, with the numerical values of $c_n$ given in (\ref{cnnum}) and the functions ${\cal W}_n(s)$ defined in (\ref{Wn}). Inserting  (\ref{DnewCI}) into (\ref{delta0}) and using (\ref{phen}), we obtain the prediction of the new CIPT:
\be\label{newCI}
\alpha_s(m_\tau^2)=  0.3198 \pm 0.0042_{\rm exp}\, ^{+0.0099}_{-0.0076}\,_{\rm K_5}\, ^{+ 0.0015}\hspace{-1cm}_{- 0.0019}\, _{\rm scale}, 
\ee
where the experimental error is due to the uncertainty quoted  in (\ref{phen}),  the second error is obtained by varying the coefficient $K_5$ given in (\ref{K5}) by  $\pm$ 50\%,  and the last error is obtained by varying the scale $\xi$ between  $1-0.63$ and  $1+0.63$ \cite{Davier2008}.

For the new FOPT  the coefficients $\tilde c_n(s)$ defined in (\ref{BwgammaFO}) as:
\bea\label{tildecnnum}
&&\hspace{-0.5cm}\tilde c_0(s)=1, \nonumber\\
&&\hspace{-0.5cm} \tilde c_1(s)= -0.7973 -2.6667 \,\eta_0, \nonumber\\
&&\hspace{-0.5cm} \tilde c_2(s)= 0.4095 - 2.4610\, \eta_0+ 3.5556\, \eta_0^2,\nonumber\\
&& \hspace{-0.5cm}\tilde c_3(s)=8.6647 - 0.6704 \,\eta_0 + 8.1489 \, \eta_0^2- 3.1605 \,\eta_0^3,\nonumber\\
&&\hspace{-0.5cm}\tilde c_4(s)=2.2416 - 18.1308 \,\eta_0 +6.7302\, \eta_0^2 - 11.0153 \, \eta_0^3,\nonumber\\&&~~~~+ 2.107 \,\eta_0^4,
\eea
where $\eta_0=\ln(s/s_0)$.

The  new FOPT  is given by the expansion (\ref{DnewFO}) truncated after 5 terms, with the numerical values of $\tilde c_n(s)$ given in (\ref{tildecnnum}) and $\tilde {\cal W}_n$ defined in (\ref{WnFO}). Inserting (\ref{DnewFO}) into  (\ref{delta0}) and using (\ref{phen}), we obtain the prediction from the new FOPT:
\be\label{newFO}
\alpha_s(m_\tau^2)=  0.3113 \pm 0.0038_{\rm exp}\,\pm 0.0013_{\rm K_5}\, ^{+ 0.0103}\hspace{-1cm}_{- 0.0006}\, _{\rm scale},
\ee
where the errors have the same significance as in (\ref{newCI}). 

As seen from  (\ref{newCI}) and  (\ref{newFO}), CIPT  is more sensitive to the uncertainty of the last perturbative term, while  FOPT  is sensitive to the ambiguity of the renormalization scale.  This  is similar to what is obtained with the standard summations \cite{Davier2008}. 

Before discussing in more detail these predictions, it is useful to investigate
the new expansions given in Section \ref{sec:new}  in the case of a physical model for the Adler function  \cite{BeJa}. 
 
\section{Model of Beneke and Jamin}\label{sec:BJ}
The physical model proposed in  \cite{BeJa} is a  parametrization of the Borel transform $B(u)$, consisting of one UV renormalon and two IR renormalons with specified branch point behaviour,  multiplied by  polynomials. The parameters of the model were  adjusted such as to reproduce the first five coefficients $K_n$ given in Eqs. (\ref{Ki}) and  (\ref{K5}). The explicit expressions and the values of the parameters are given in Section 6 of   \cite{BeJa}, and we shall not reproduce them here. 

The new CIPT and FOPT can be constructed in a staightforward way:  from the  parameters $K_n$ of the model\footnote{The values from $K_1$ to $K_5$ are given in  Eqs. (\ref{Ki}) and  (\ref{K5}); the next 6 values are given in Table 2 of \cite{BeJa}: $K_6=3275$, $K_{7}=1.88 \cdot 10^{4}$, $K_{8}=3.88 \cdot 10^{5}$,   $K_{9}=9.19 \cdot 10^{5}$, $K_{10}=8.37 \cdot 10^{7}$, $K_{11}=-5.19 \cdot 10^{8}$, $K_{12}=3.38 \cdot 10^{10}$; for completeness, we list the next 5 parameters: $K_{13}=-6.04 \cdot 10^{11},\, K_{14}=2.34 \cdot 10^{13},\, K_{15}=-6.52 \cdot 10^{14},\,  K_{16}=2.42 \cdot 10^{17},\,  K_{17}=-8.46 \cdot 10^{17},\,  K_{18}=3.36 \cdot 10^{19}$.} one computes the Borel function (\ref{B}) truncated after a certain number of terms $N$. The coefficients $c_n$ and $\tilde c_n(s)$ at that order are calculated from the expansions   (\ref{Bwgamma}) and (\ref{BwgammaFO}), respectively. The new  expansions are given by (\ref{DnewCI}) and (\ref{DnewFO}), truncated at the same number of terms $N$, the expansion functions  being defined in (\ref{Wn}) and  (\ref{WnFO}), respectively.

As in \cite{BeJa}, we assume that the expansion of the $\beta$ function contains only four terms, with the  coefficients given by (\ref{betai}). For  $\alpha_s(m_\tau^2)=0.34$ the exact value of $\delta^{(0)}$, obtained with the PV of the Borel sum, Eq. (\ref{Dpv}), is  \cite{BeJa}
\be\label{BS}
\delta^{(0)}_{BS}=0.2371 \pm 0.0060\, i,
\ee
where the error is an estimate of the  prescription ambiguity  (cf. Eq. (6.3) of  \cite{BeJa}). We note that the "exact" results are obtained  by using the contour improved Borel sum (\ref{Laplace}), with $B(u)$ of the form specified by the model.

\begin{figure}\begin{center}
\includegraphics[width=6.5cm]{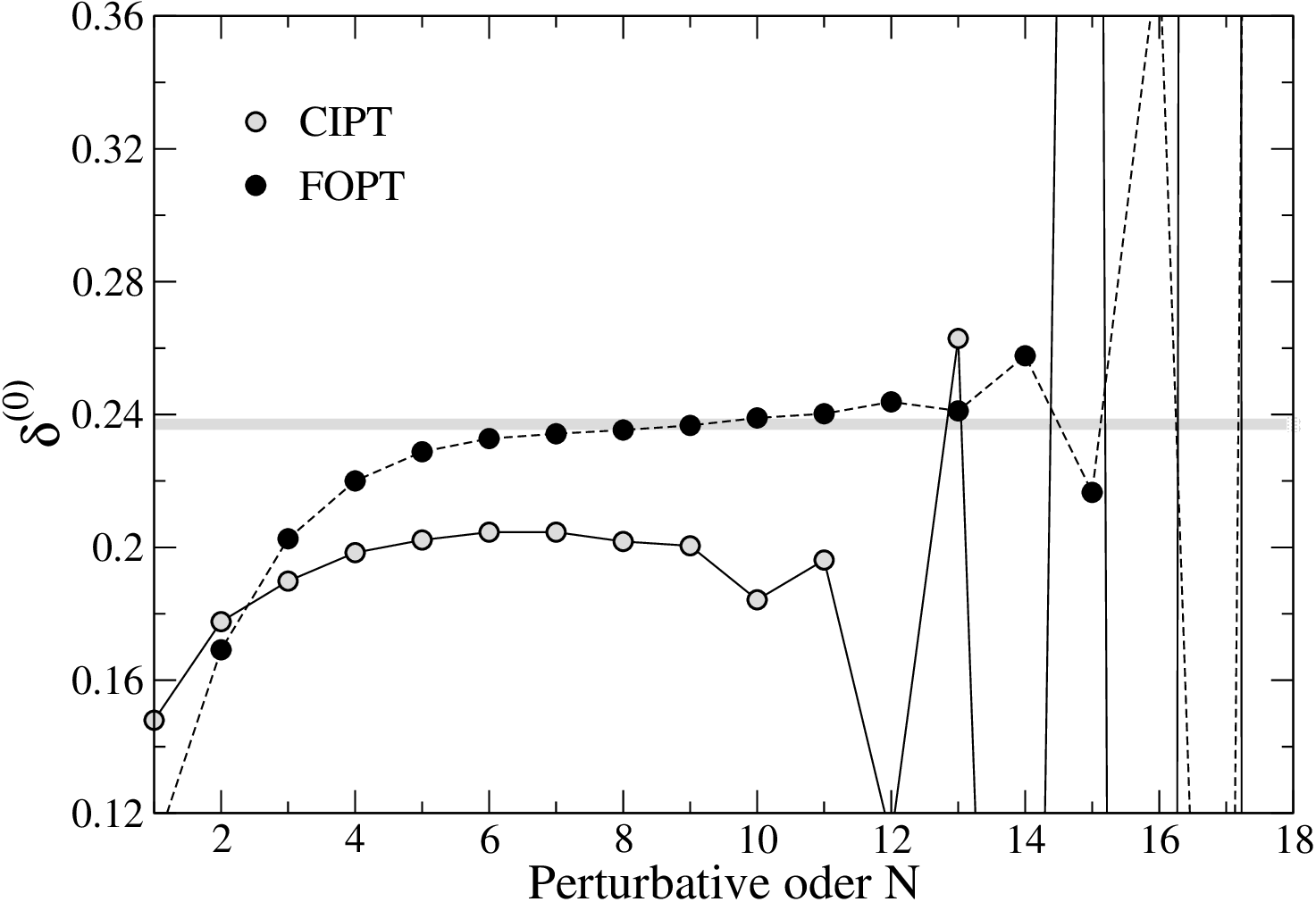}
\caption{\label{fig:CIFO} Values of $\delta^{(0)}$ for the model of Beneke and Jamin  calculated with the standard CIPT and FOPT, as a function of the order up to which the series have been summed. The horizontal band is the exact value (\ref{BS}). }\end{center}
\end{figure}	
\begin{figure}\begin{center}
\includegraphics[width=6.5cm]{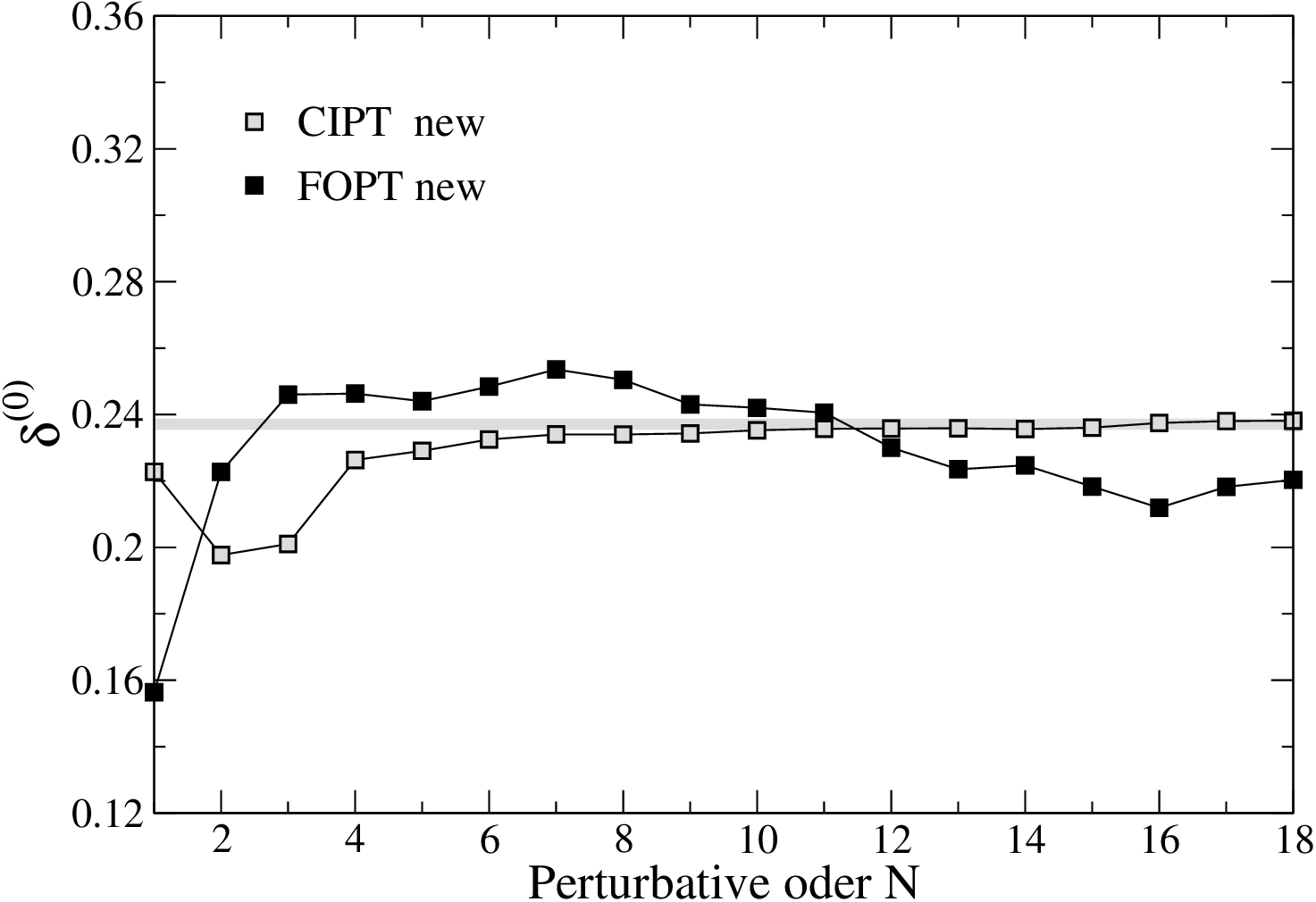}
\caption{\label{fig:CIFOw} Values of $\delta^{(0)}$ for the model of Beneke and Jamin, calculated with the new CIPT and FOPT defined by Eqs. (\ref{DnewCI}) and (\ref{DnewFO}), respectively,  as a function of the perturbative order $N$.   The horizontal band is the exact value (\ref{BS}). }\end{center}
\end{figure}	

The comparison of the standard CIPT and FOPT with the new CIPT and FOPT is seen from  Figs. \ref{fig:CIFO} and    \ref{fig:CIFOw}, where we show  $\delta^{(0)}$ calculated as a function of the order up to which the series have been summed.  Fig. \ref{fig:CIFO} reproduces Fig. 7 of \cite{BeJa}, and shows that the standard CIPT does not approach the true value, staying below it up to the orders at which the results start to exhibit large oscillations. On the contrary, as seen in  Fig.  \ref{fig:CIFOw}, the new CIPT gives very good results which approach the true value with great accuracy when $N$ increases. As concerns FOPT, the new approach gives results somewhat poorer than the standard one at low orders. At large orders, when the standard FOPT shows large oscillations, the new FOPT leads to values closer to the true result, but not as good as those obtained with the new CIPT.  

\begin{figure}\begin{center}
\includegraphics[width=6.1cm]{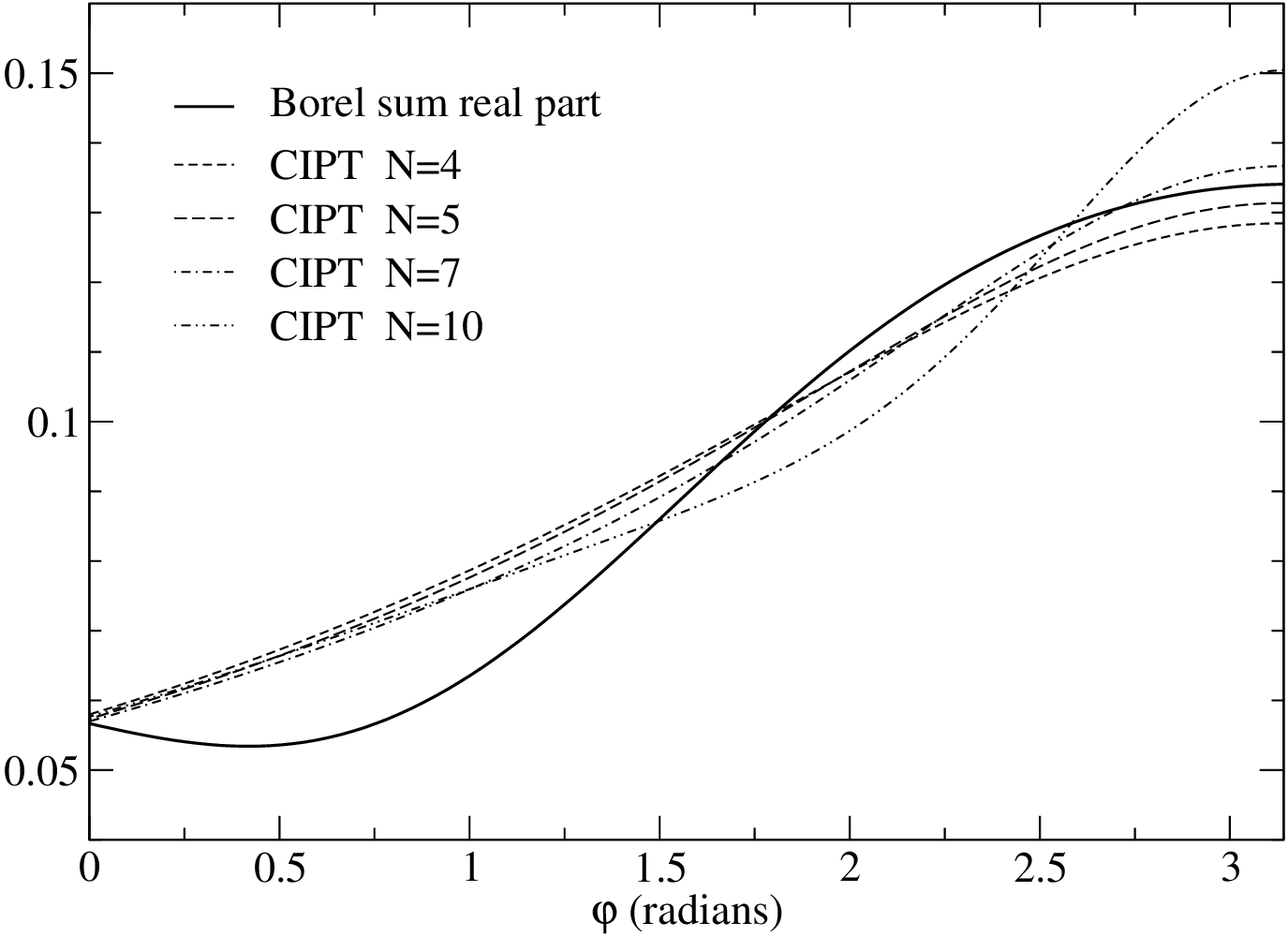}
\caption{\label{fig:DRCI} Real part of $\hat D(s)$ for $s= m_\tau^2 \rm{e}^{i \varphi}$, calculated  with the standard CIPT, Eq. (\ref{DCI}), for various numbers $N$  of perturbative terms. }\end{center}
\end{figure}

\begin{figure}\begin{center}
\includegraphics[width=6.1cm]{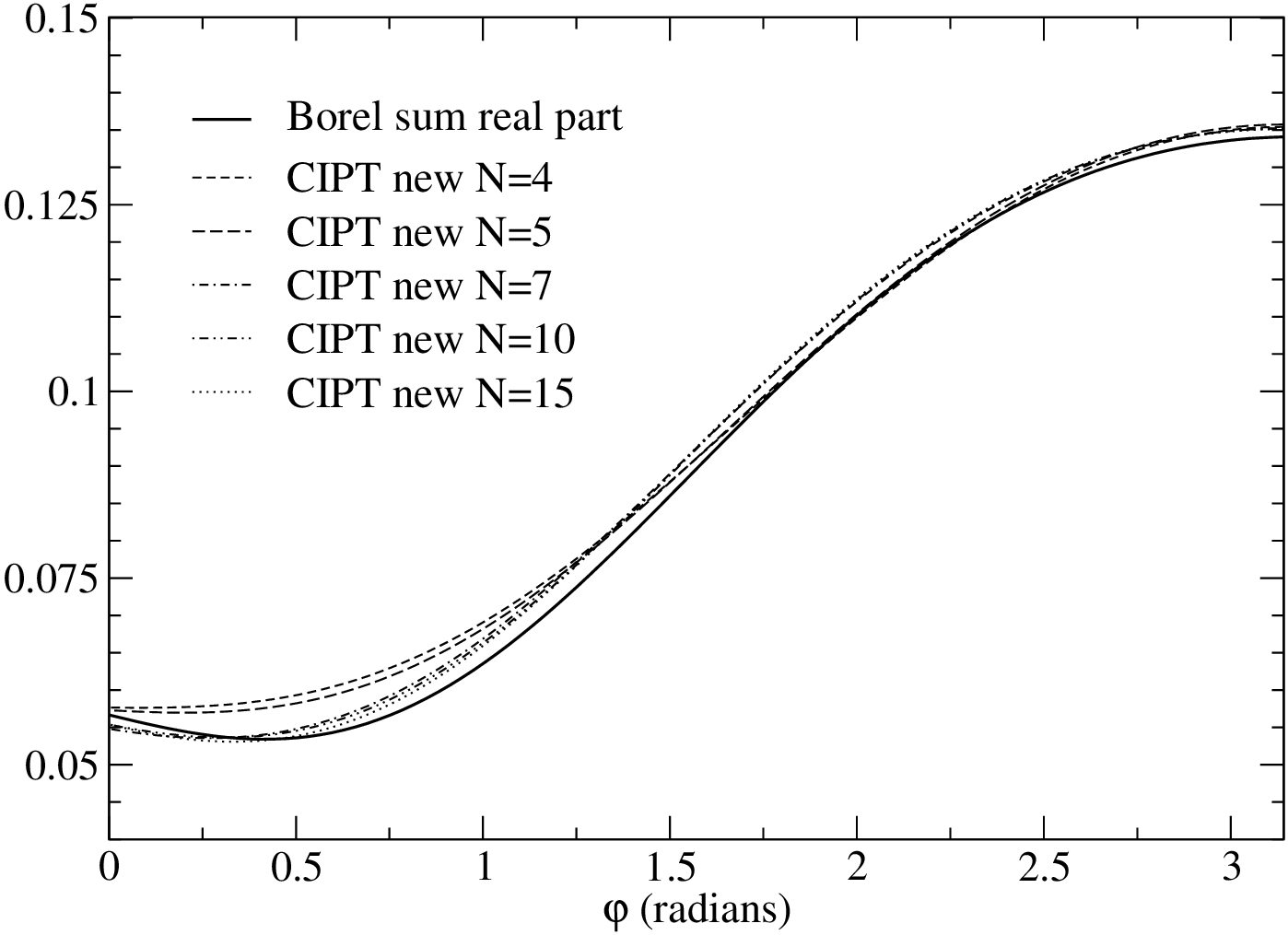}
\caption{\label{fig:DRCIw} Real part of $\hat D(s)$ for $s= m_\tau^2\, {\rm e}^{i \varphi}$, calculated with the new CIPT, Eq. (\ref{DnewCI}).}\end{center}
\end{figure}

\begin{figure}\begin{center}
\includegraphics[width=6.1cm]{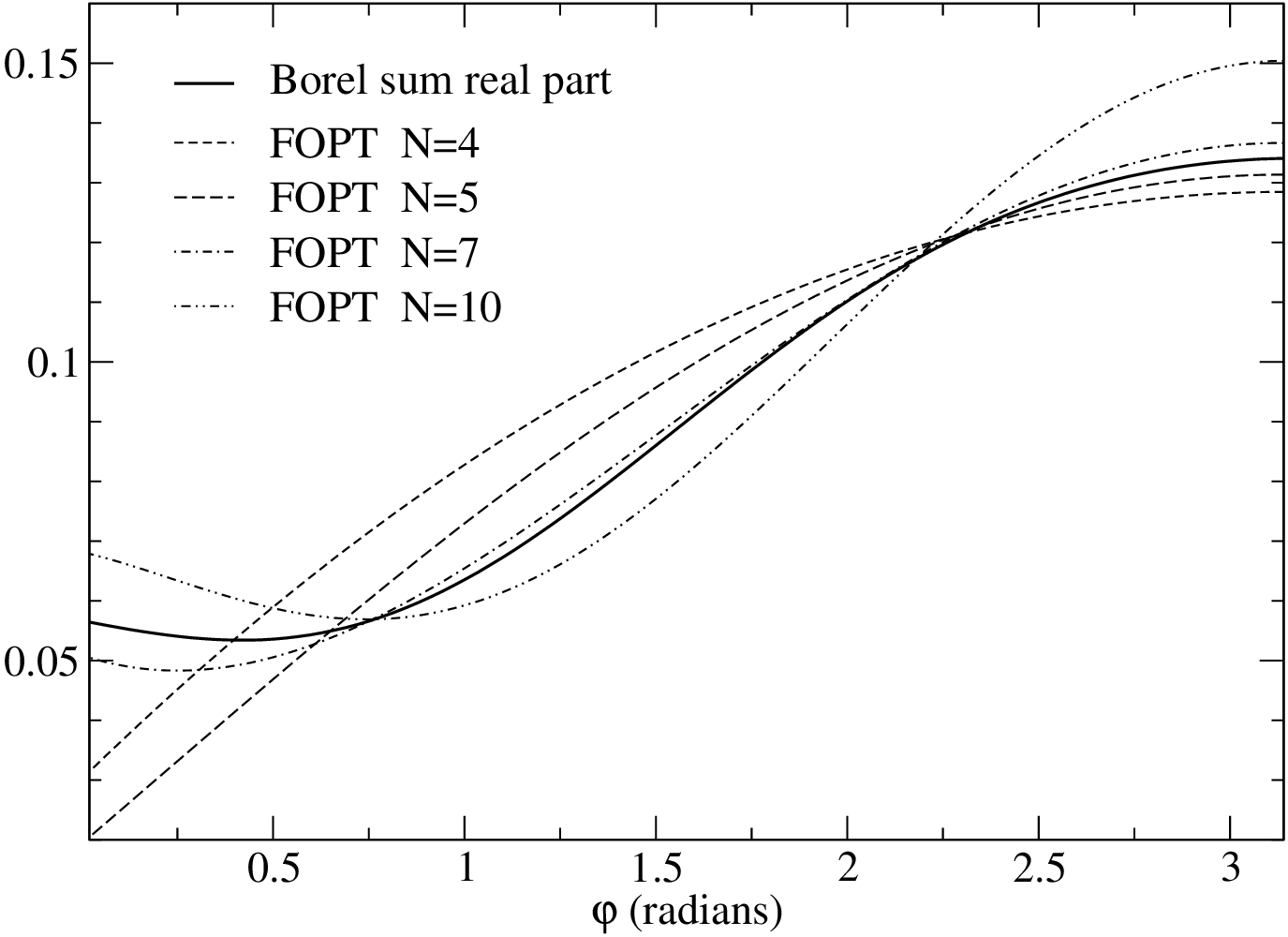}
\caption{\label{fig:DRFO}  Real part of $\hat D(s)$ for $s= m_\tau^2 \rm{e}^{i \varphi}$ calculated  with the standard FOPT, Eq. (\ref{DFO}). }\end{center}
\end{figure}

\begin{figure}\begin{center}
\includegraphics[width=6.1cm]{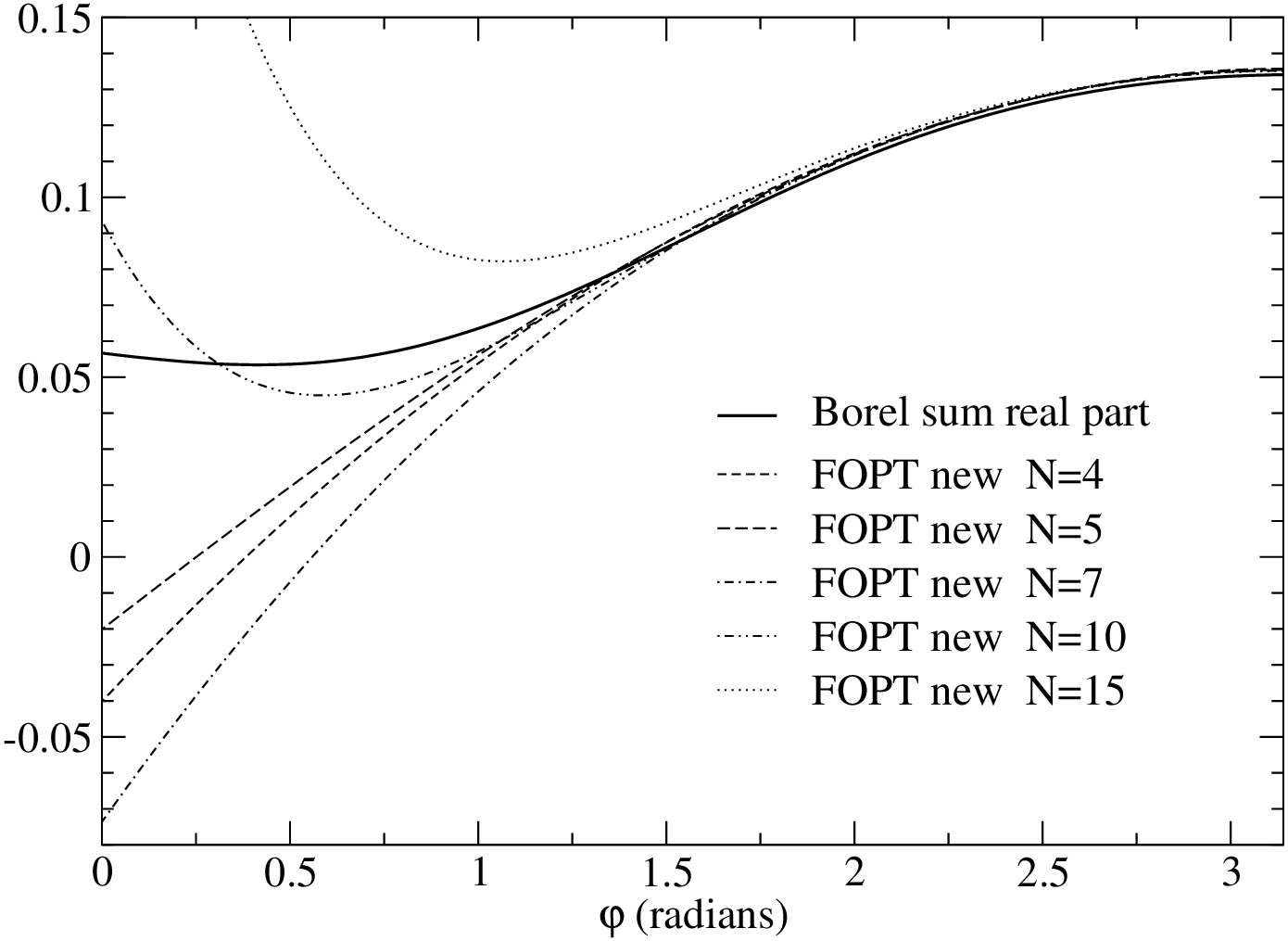}
\caption{\label{fig:DRFOw} Real part of $\hat D(s)$ for $s= m_\tau^2 \rm{e}^{i \varphi}$ calculated with the new FOPT, Eq. (\ref{DnewFO}). }\end{center}
\end{figure}

\begin{figure}\begin{center}
\includegraphics[width=6.1cm]{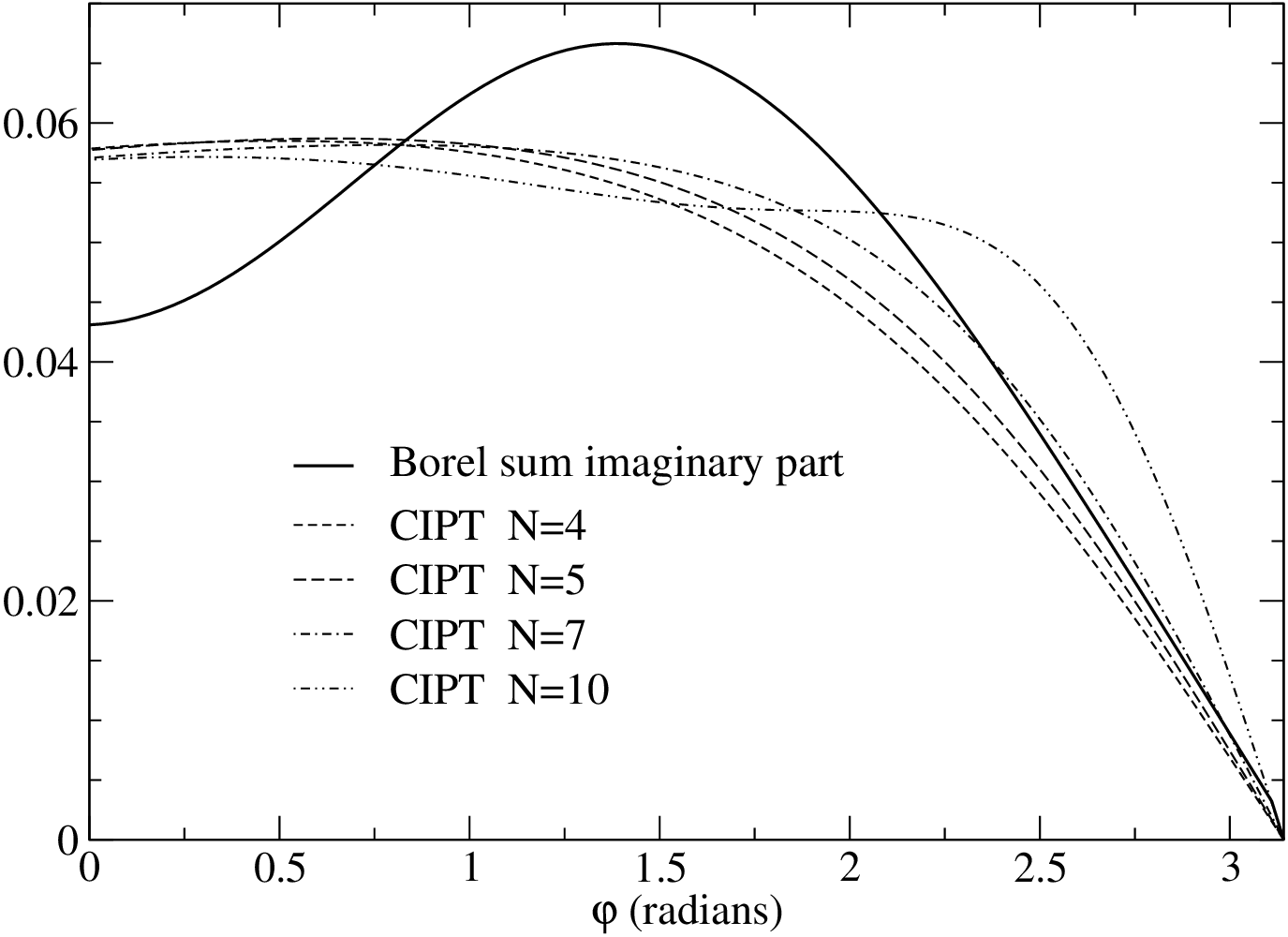}
\caption{\label{fig:DICI} Imaginary part of $\hat D(s)$ for $s= m_\tau^2 \rm{e}^{i \varphi}$ calculated with the standard CIPT, Eq. (\ref{DCI}). }\end{center}
\end{figure}

\begin{figure}\begin{center}
\includegraphics[width=6.1cm]{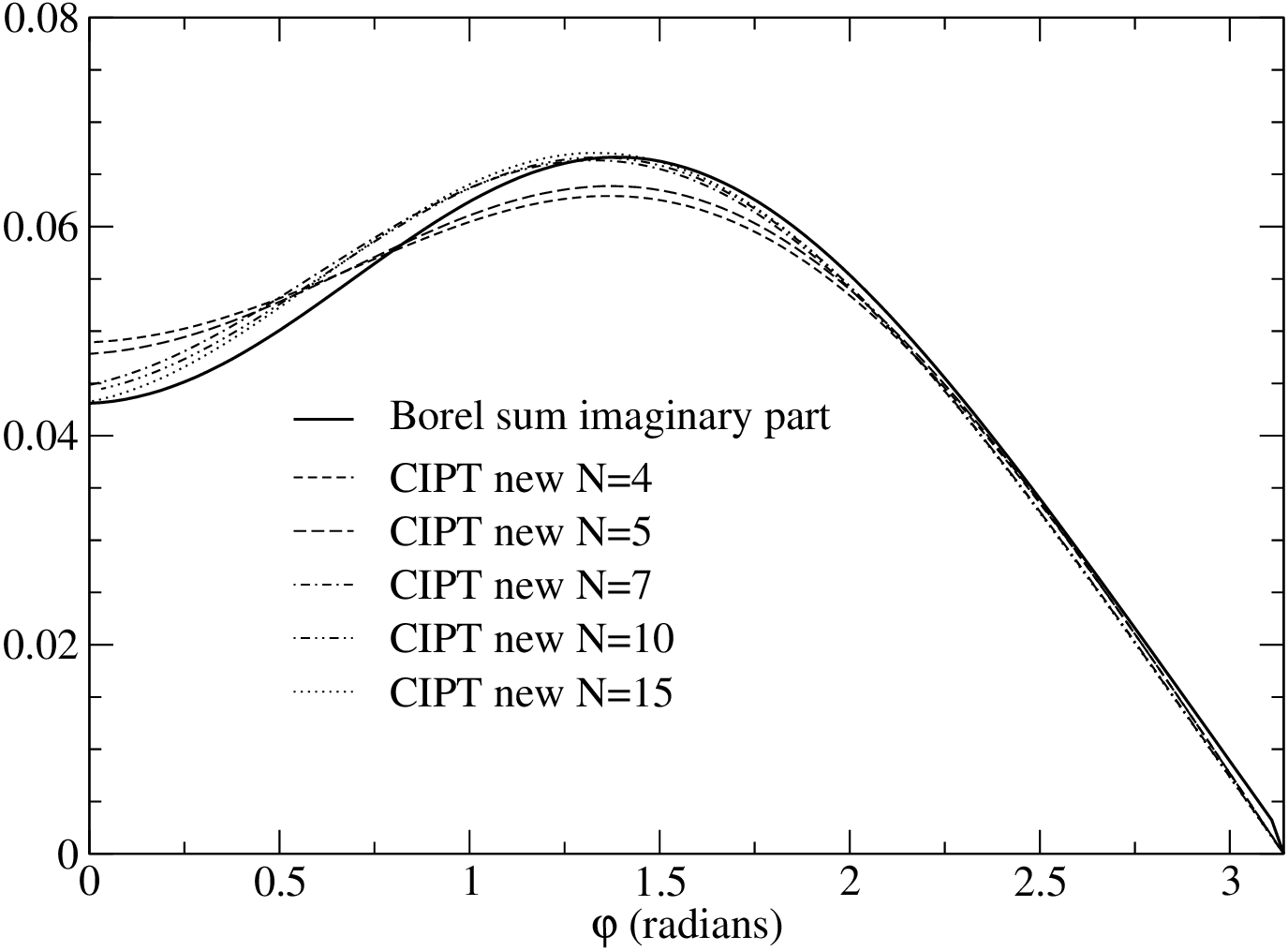}
\caption{\label{fig:DICIw} Imaginary part of  $\hat D(s)$ for $s= m_\tau^2 \rm{e}^{i \varphi}$ calculated with the new CIPT, Eq. (\ref{DnewCI}). }\end{center}
\end{figure}

\begin{figure}\begin{center}
\includegraphics[width=6.1cm]{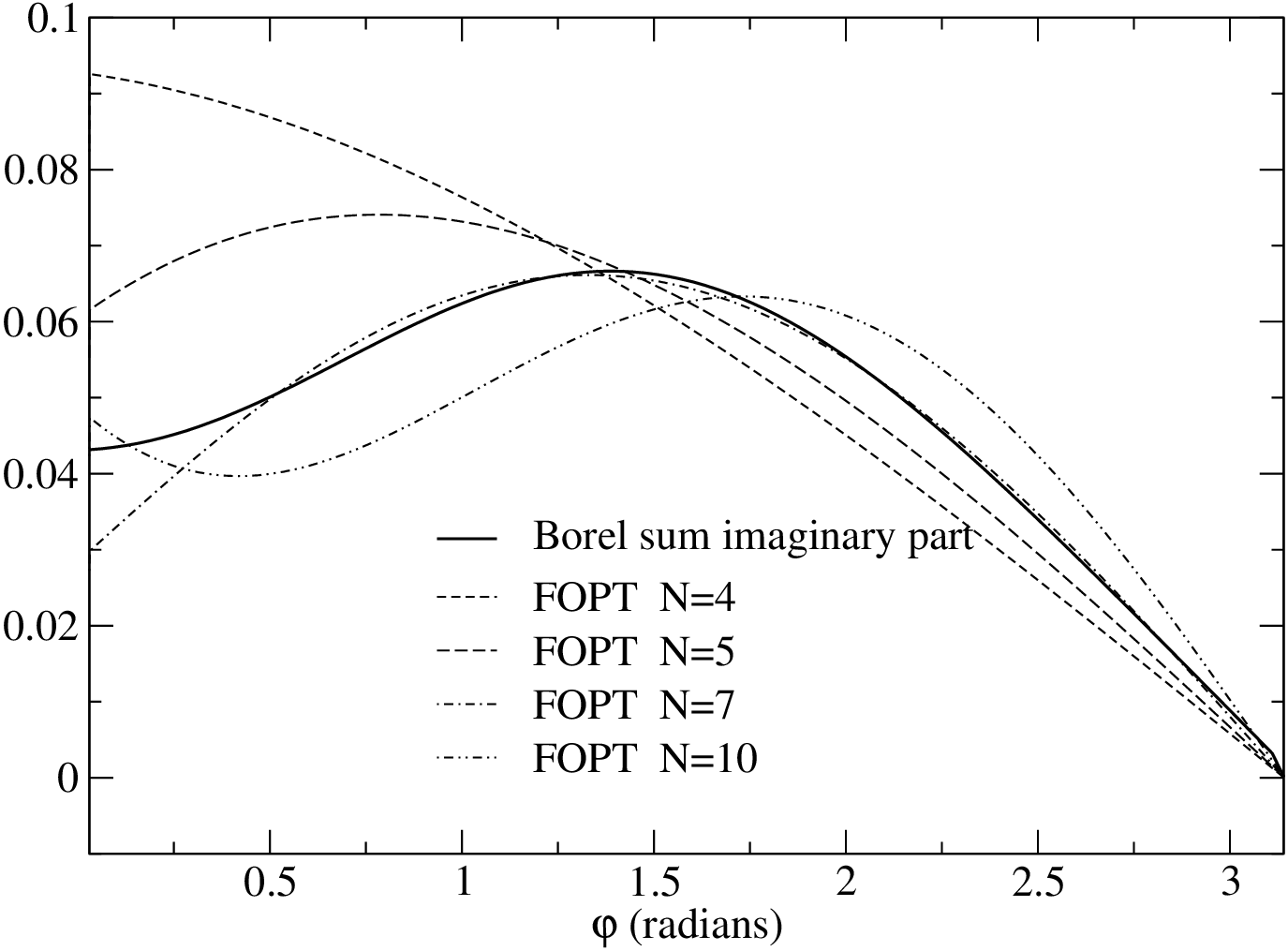}
\caption{\label{fig:DIFO} Imaginary part of $\hat D(s)$ for $s= m_\tau^2 \rm{e}^{i \varphi}$ calculated with the standard FOPT, Eq. (\ref{DFO}). }\end{center}
\end{figure}

\begin{figure}\begin{center}
\includegraphics[width=6.1cm]{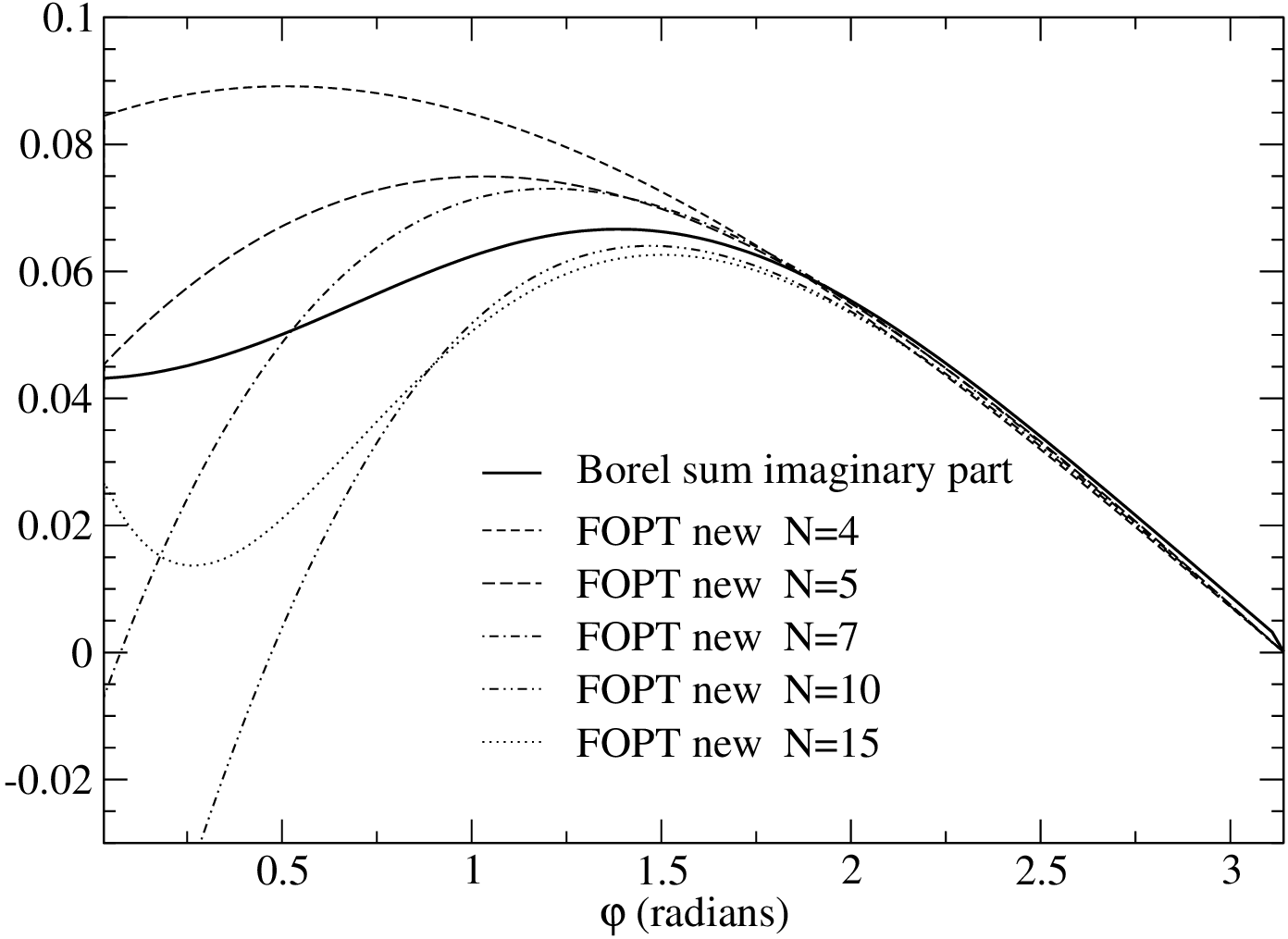}
\caption{\label{fig:DIFOw} Imaginary part of  $\hat D(s)$ for $s= m_\tau^2 \rm{e}^{i \varphi}$ calculated with the new FOPT,  Eq. (\ref{DnewFO}).}\end{center}
\end{figure}

In order to understand the origin of these results, we calculated the Adler function $\hat D(s)$ for complex $s$ along the integration contour. In Figs. \ref{fig:DRCI} - \ref{fig:DRFOw}  we present the real part of  $\hat D(s)$ calculated with the standard/new CIPT and FOPT, for $s$  along the upper semicircle in the definition (\ref{delta0}) of $\delta^{(0)}$ ($s= m_\tau^2 \rm{e}^{i \varphi}$, for $0\le \varphi \le \pi$). Figs. \ref{fig:DRCI} and \ref{fig:DRFO} reproduce Fig. 9 of \cite{BeJa}.  In Figs. \ref{fig:DICI} - \ref{fig:DIFOw} we present the imaginary part of $\hat D(s)$ for the same values of $s$. Note that the values along the lower semicircle follow  immediately from the reality condition $\hat D(s^*)=
(\hat D(s))^*$.
By comparing Fig. \ref{fig:DRCI} with Fig. \ref{fig:DRCIw}, one can see that, for the new CIPT,  the quality of the approximation of the real part of  $\hat D(s)$ improves continously with increasing $N$ along the whole contour. By contrast, for the standard CIPT the low orders are not able to give a good approximation, while starting from N=10 the deviations increase dramatically (we can not show the curve for $N=15$ due to these huge oscillations).



As concerns FOPT, the comparison of Fig. \ref{fig:DRFO} with Fig. \ref{fig:DRFOw} shows that the new FOPT gives a very good  approximation to the real part of  $\hat D(s)$,  which improves continously with increasing $N$, for  $\varphi$ close to $\pi$, {\em i.e.} near the spacelike axis. However, the description deteriorates as  $\varphi$ approaches 0,  {\em i.e.} near the timelike axis.  This can be understood by the imaginary logarithms present in the  expansion (\ref{astaylor}) of the running coupling, which are large here. By contrast, the approximation provided by the standard FOPT does not reach the same precision for   $\varphi$ close to $\pi$, {\em i.e} near  the euclidian axis.

Similar conclusions are obtained for the imaginary part of  $\hat D(s)$, shown in Figs.  \ref{fig:DICI} - \ref{fig:DIFOw}: the new CIPT provides for all  $\varphi$ a very good approximation, which improves continously as $N$ increases, while the new FOPT reproduces well the imaginary part of $\hat D(s)$ in the region where the effect of the large imaginary logarithms is small. By contrast, the standard CIPT and FOPT are not able to approximate accurately the function at low $N$, and start to oscillate violently at large $N$.   In spite of the rather poor local accuracy,  the standard  FOPT at low orders gives acceptable results for $\delta^{(0)}$ because the region near $\varphi=0$ is suppressed by the factor $\omega(s)$.

\begin{figure}\begin{center}
\includegraphics[width=6.1cm]{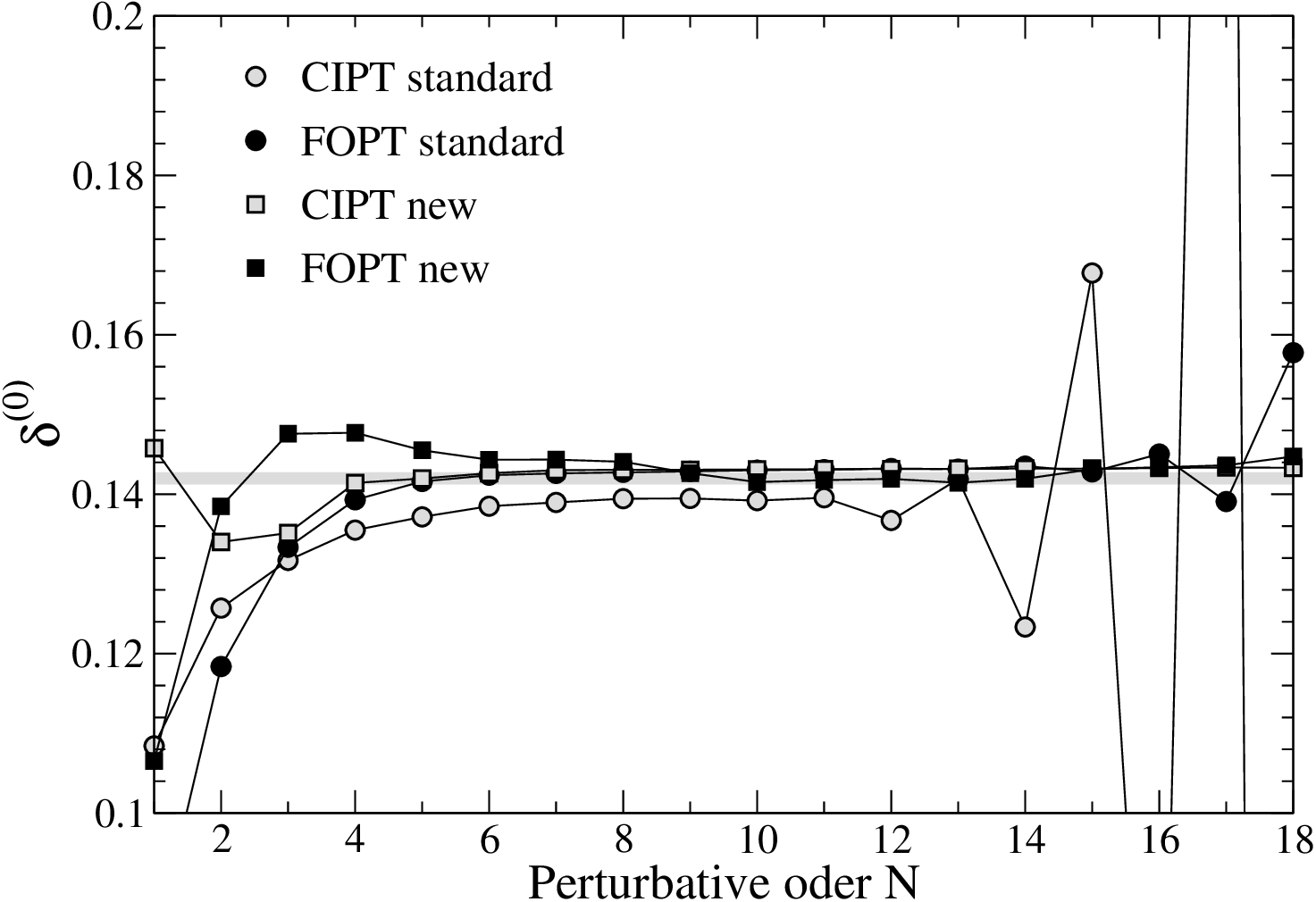}
\caption{\label{fig:CIFOw_a1}  Values of $\delta^{(0)}$ for the model of Beneke and Jamin calculated with the CI and FO expansions,  for $\alpha_s(m_\tau^2)=0.26$. }\end{center}
\end{figure}

 We recall that  the CI summation  was proposed in \cite{Pivo}, \cite{DiPi} in order to avoid the large imaginary logarithms, responsible for a slow convergence of the expansion (\ref{astaylor}) along the integration countour.  This slow convergence  affects also the new FO expansion, which has as starting point the standard expansion (\ref{DFO}). This explains the poor approximation of the Adler function by the new FO expansion at points far from the euclidian axis.  One expects of course an improved convergence  for a smaller value of the coupling.  This is confirmed by  Fig. \ref{fig:CIFOw_a1}, where we present the values of $\delta^{(0)}$  calculated for $\alpha_s(m_\tau^2)=0.26$ with the standard and new expansions. For this coupling the new CI and FO expansions are very close up to the 18th order, while the standard expansions exhibit a larger difference (although smaller than in the case of the physical value of  $\alpha_s(m_\tau^2)$).
  
\section{Discussion}\label{sec:disc}
In this Section we  make several more remarks on the expansions in powers of conformal mappings  and their implications for the determination of $\alpha_s$. As discussed in Section \ref{sec:new}, the definition of the optimal variable (\ref{w}) requires only the knowledge of the location of the  singularities of $B(u)$  in the Borel plane. The most general  expansions based on the powers of the optimal variable are (\ref{DnewCI0}) and  (\ref{DnewFO0}) for the CI and FO summations, respectively.

\begin{figure}\begin{center}
\includegraphics[width=6.5cm]{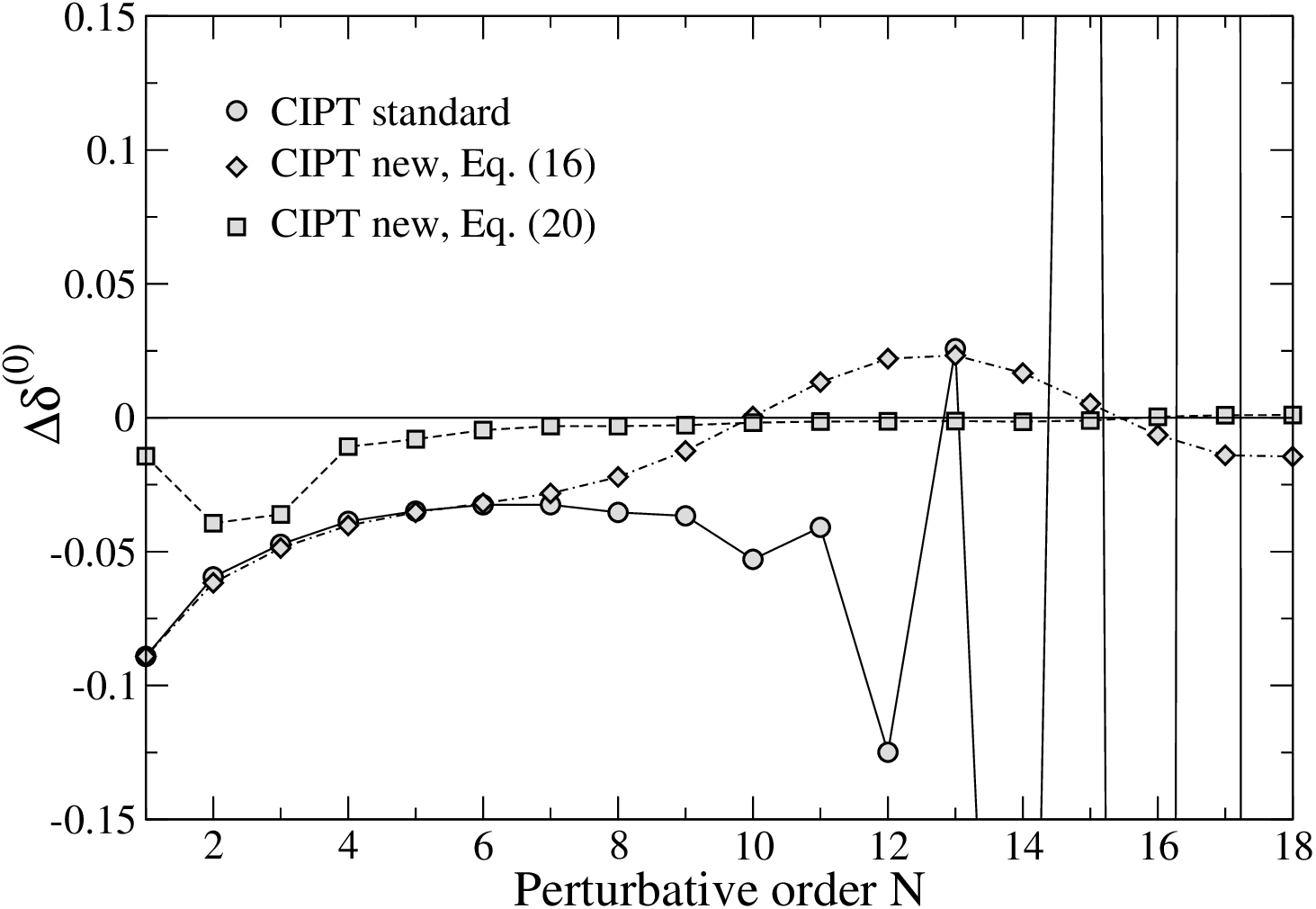}
\caption{\label{fig:CIerrww0} Error of the determination of $\delta^{(0)}$ as a function of perturbative  order, for the standard and the new CI expansions.} \end{center}
\end{figure}

\begin{figure}\begin{center}
\includegraphics[width=6.5cm]{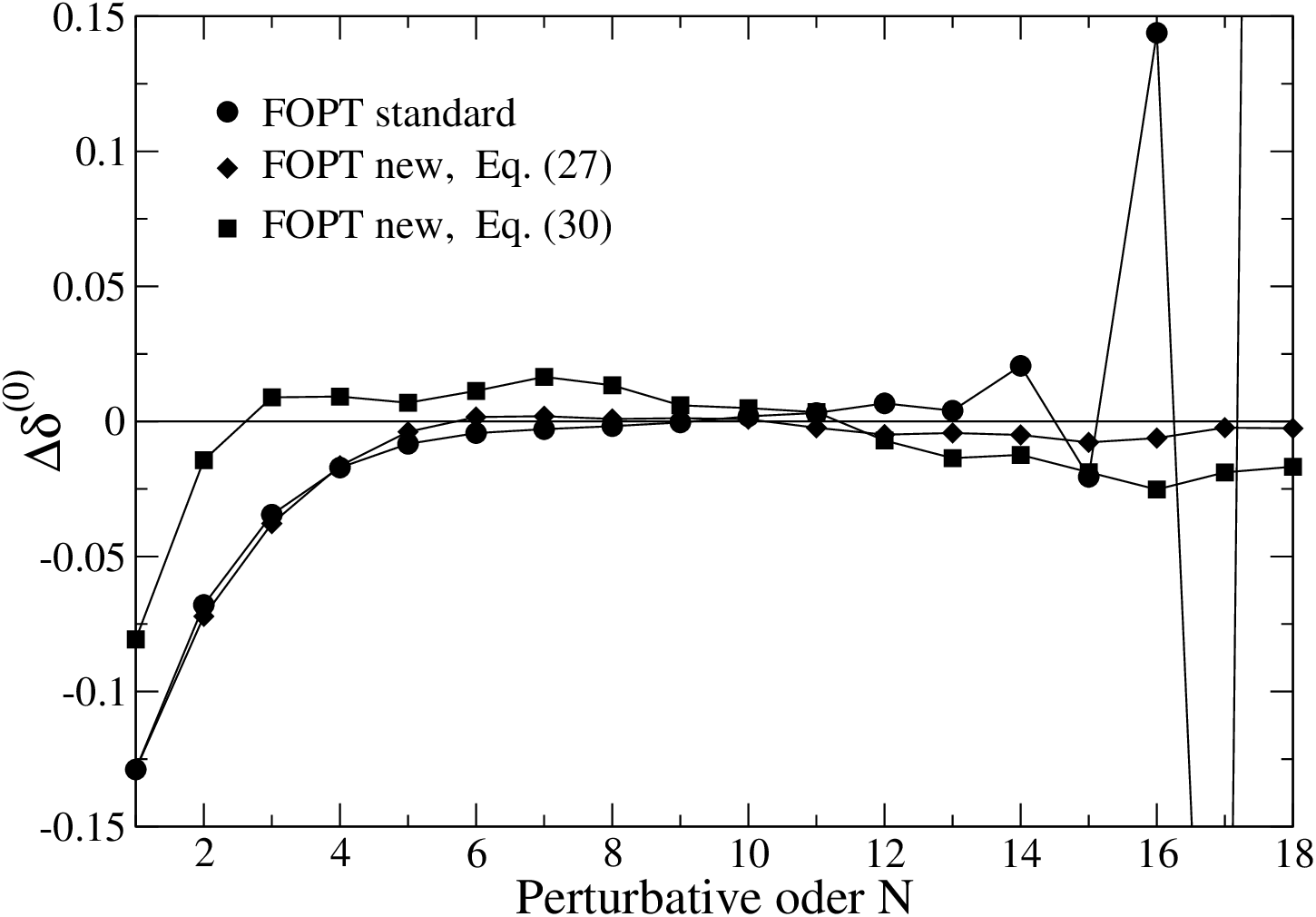}
\caption{\label{fig:FOerrww0} Error of the determination of $\delta^{(0)}$ as a function of perturbative  order, for the standard  and the new FO expansions.} \end{center}
\end{figure}

In the analysis presented in the previous Sections we used the modified expansions  (\ref{DnewCI}) and (\ref{DnewFO}), generated by the Borel series (\ref{Bwgamma}) and  (\ref{BwgammaFO}) respectively,   which explicitly include
 the nature of the lowest (leading) renormalon singularities in the Borel plane. It is of interest to consider also  the general expansions (\ref{DnewCI0}) and  (\ref{DnewFO0}), whose respective counterparts (\ref{Bw0}) and  (\ref{BwFO})   do  not manifestly display this singular behaviour.  According to general arguments \cite{CiFi}, we expect them to approach the true result, if the number of the perturbative terms is large enough. This is confirmed in Figs. \ref{fig:CIerrww0} and  \ref{fig:FOerrww0}, where  we show the error of the determination of $\delta^{(0)}$ as a function of perturbative order $N$, for the standard CIPT and FOPT and the new expansions defined above. 

 For small $N$, the expansions (\ref{DnewCI0}) and (\ref{DnewFO0})  give results  similar to the standard CI and FO expansions, while the  expansions (\ref{DnewCI}) and (\ref{DnewFO}) approach much better the true result. This shows that at low $N$ the most important effect is the factorization of the singularities.   However, for $N$ greater than 7 the effect of the conformal mapping becomes visible: as illustrated in  Fig. \ref{fig:CIerrww0},  while the standard CI expansion starts to show large oscillations, the  expansion  (\ref{DnewCI0}) based on the conformal mapping of the complex Borel plane leads to only small oscillations around the true value. The improvement brought by the conformal mapping  becomes visible rather slowly  due to the strong leading IR singularity of the physical  model considered in \cite{BeJa}.

 For the FO summation,  the standard expansion  starts to oscillate wildly for $N\ge 14$, while both  modified expansions  remain close to the true result. 
In this case,  the approximation provided by the expansion (\ref{DnewFO0}), which does not include the nature of the singularities, turns out to be slightly
 better than the modified expansion (\ref{DnewFO}). As discussed in Section \ref{sec:BJ}, all the FO  expansions provide a less satisfactory description of the Adler function  near the timelike axis, but  this region is suppressed in the integral (\ref{delta0}).


The good approximations provided by the new CIPT (and, at points where the  convergence of the series (\ref{astaylor}) is not poor, also by the new FOPT) can be explained qualitatively
 by a theorem proved  in \cite{CiFi}, which implies that the expansion (\ref{Bw0}) in powers of the variable $w$ has a better rate of convergence than the standard expansion (\ref{B}) at points of the complex Borel plane where both converge. In particular, the convergence is better at points on the real positive axis below the first IR singularity, which gives the dominant  contribution to  the  Laplace-Borel integral (\ref{Laplace}), due to the exponential factor
 that  strongly suppresses the contribution of  large $u$. Therefore, the improved expansion of $B(u)$  leads to a better approximation of the integral, at least for the perturbative orders investigated up to now.  One might of course ask what happens if  $N$ is further increased.  It can be shown that going  up to the 36th order the new  CIPT continues to be stable, while the new FOPT somewhat deteriorates, and this is valid also for modified models, having, for instance, a smaller residue of the leading IR renormalon, compensated by an additional IR singularity that preserve the known low-order coefficients\footnote{We thank M. Jamin for communicating to us these results.}. However,  a divergent behaviour at still higher orders is not excluded in principle, taking into account that the Laplace-Borel integral is performed along the cut, while the series (\ref{Bw0}) converges only at interior points. For a discussion of this problem and  explicit criteria of convergence of the new expansion in the one-loop approximation of the coupling  we refer to \cite{CaFi1}.

As mentioned in Section \ref{sec:new}, the  inclusion  of the behaviour of $B(u)$ at the first branch points is not unique.  Of course, for a large number of terms in the expansion the form of these factors is irrelevant, but at low orders one prescription may be better than the other.  In (\ref{DnewCI}) and (\ref{DnewFO}),  the dominant singular factors were expressed in the $w$ variable. Other  possible expansions  will be investigated  elsewhere.

In order to test the choice made in this work, we use an idea applied  in \cite{CaFi} for illustrating  the effect of conformal mappings on  the convergence of power series. Consider the expansion (\ref{Bwgamma}) truncated after $N$ terms. The coefficients $c_n$, $n\le N-1$, are calculated in a straightforward way  from the first $N$ coefficients $K_n$, assumed to be known.  When expanded as a Taylor series in the variable $u$, this expression reproduces  the  $N$ coefficients  $K_n$ used as input, but contains also an infinity of higher order terms, in particular it predicts the next coefficient $K_{N+1}$. As an exercise, we used the expansion (\ref{Bwgamma}) with 4 terms, using as input the coefficients $K_1$ to $K_4$. The coefficients $c_0$ to $c_3$ are given in (\ref{cnnum}), the coefficients $c_n$ with $n\ge 4$  are set to zero. By reexpressing $B(u)$ from (\ref{Bwgamma}) as a series in powers of the $u$ variable, and using (\ref{B}) and (\ref{bn}),  we obtain $K_5=256$, which is quite close to the value $K_5=283$  adopted as a good estimate in \cite{BeJa}.  So, the expansion  (\ref{Bwgamma}) is able to reproduce to a reasonable extent the higher order coefficients of the expanded function.

 This nice feature is even more striking  if we go one  step further,  using as input  5 coefficients $K_n$  from (\ref{Ki}) and   (\ref{K5}). Then we have 5  nonzero coefficients $c_n$, given in (\ref{cnnum}). By expanding $B(u)$ from (\ref{Bwgamma}) as a series in powers of  $u$, we predict the higher coefficients $K_6=2929$, $K_7=1.73 \cdot 10^4$, $K_8=3.14 \cdot 10^5$,   $K_9=9.23 \cdot 10^5$,   $K_{10} =
6.34 \cdot 10^7$,    $K_{11} = -3.52 \cdot 10^8$,   $K_{12} = 2.48 \cdot 10^{10}$, which agree qualitatively with the values of the model of Beneke and Jamin \cite{BeJa}, given in  footnote 4. These apparently miraculous predictions are explained by the fact that the new expansion incorporates some features of the expanded function, which are known in advance. Then,   with a smaller number of terms in the variable $w$, it reproduces well the higher order coefficients in the variable $u$.

\begin{figure}\begin{center}
\includegraphics[width=6.1cm]{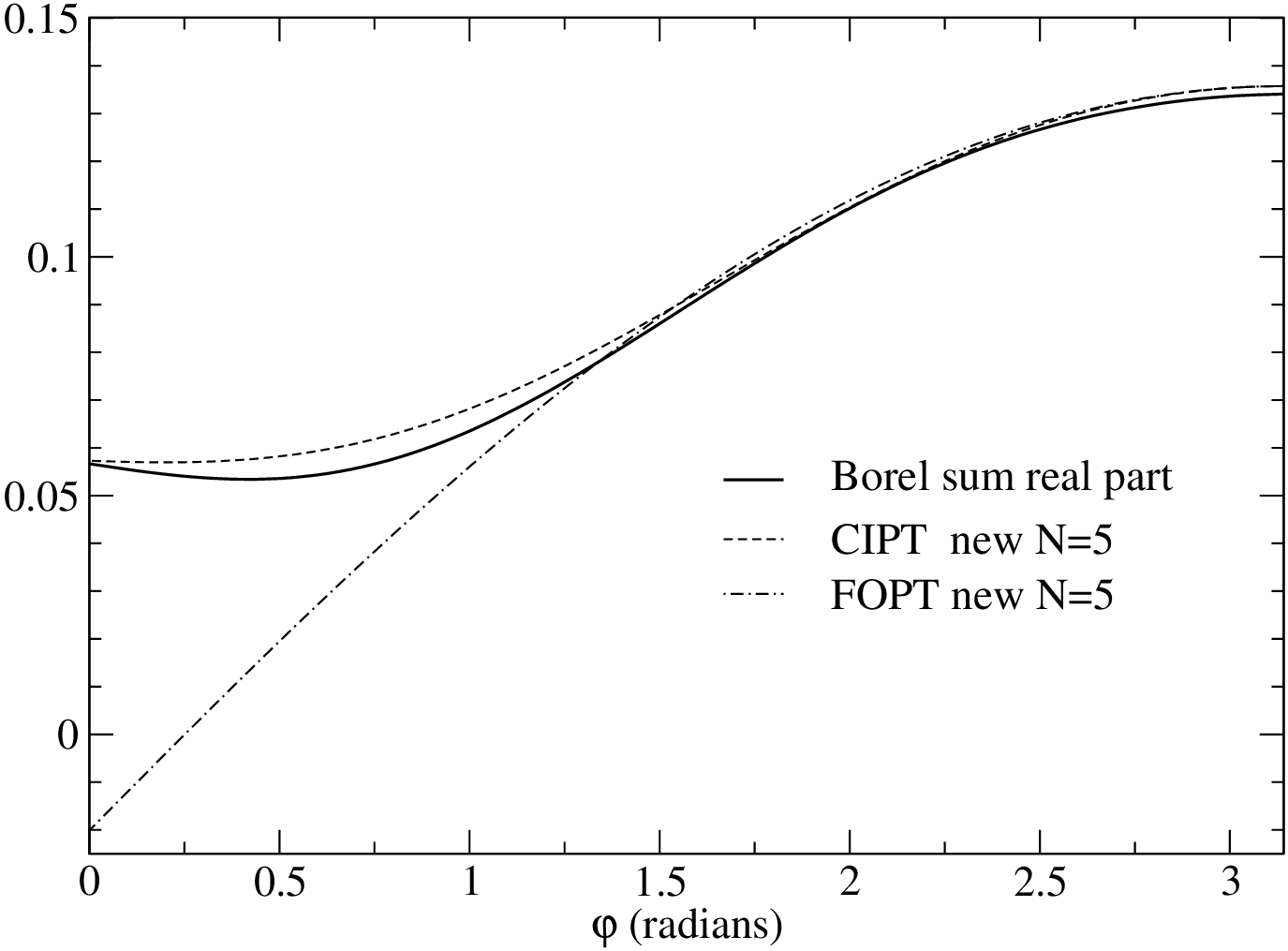}
\caption{\label{fig:DRCIFO5} Real part of $\hat D(s)$ for $s= m_\tau^2 \rm{e}^{i \varphi}$ calculated with five perturbative terms in the new CIPT, Eq.(\ref{DnewCI}), and new FOPT, Eq.(\ref{DnewFO}). }\end{center}
\end{figure}	
\begin{figure}\begin{center}
\includegraphics[width=6.1cm]{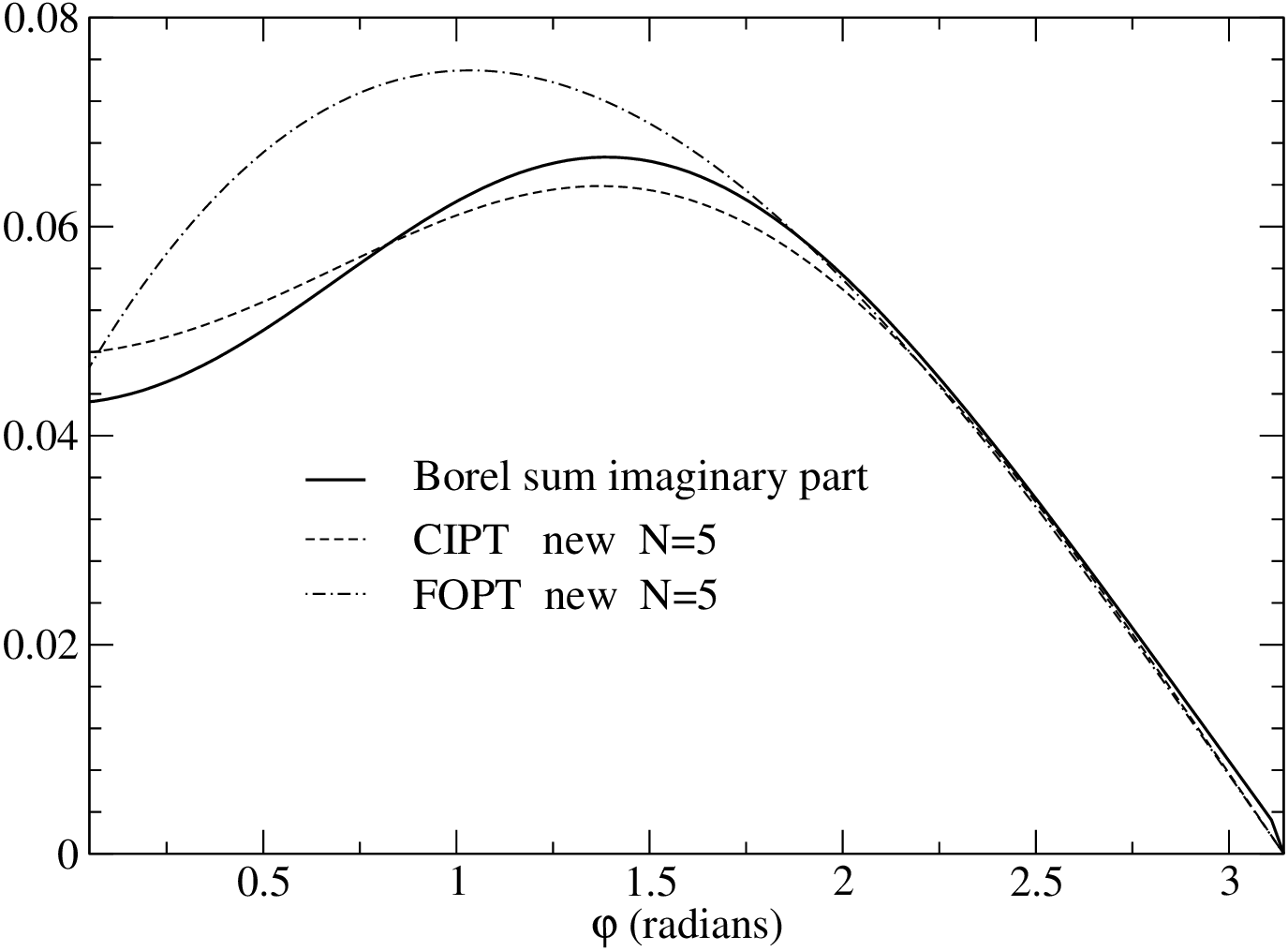}
\caption{\label{fig:DICIFO5} Imaginary part of $\hat D(s)$ for $s= m_\tau^2 \rm{e}^{i \varphi}$ calculated with five perturbative terms in the new CIPT, Eq.(\ref{DnewCI}), and new FOPT, Eq.(\ref{DnewFO}).}\end{center}
\end{figure}	
Finally, let us consider in more detail the predictions for $N=5$, which is the number of terms known in the physical case (recall that we adopted the value of $K_5$ used  in  \cite{BeJa}).  As seen from Fig. \ref{fig:CIFOw}, for $N=5$ the new CIPT and FOPT give results which approximate well the true value of $\delta^{(0)}$  from above and below, respectively. These results can be understood from the plots of the real and imaginary part of  $\hat D(s)$  shown in Figs. \ref{fig:DRCIFO5}-\ref{fig:DICIFO5}. The new CIPT and FOPT  approximate  very well the true functions for large $\varphi$. For intermediate values of  $\varphi$, CIPT and FOPT approach the true values comparatively well from opposite sides. For  $\varphi$    close to 0, CIPT gives definitely a better approximation, but this region is suppressed in the integral (\ref{delta0}). This explains why the resulting values of $\delta^{(0)}$ given by the new CIPT and FOPT are comparable.
	
\section{Summary and conclusions}\label{sec:conc}
In this paper we applied a new perturbation series for QCD observables, proposed in \cite{CaFi}, to the two renormalization group summations, CI and FO, used for  the extraction of  $\alpha_s$ from $\tau$ decays. The new expansion, which replaces the standard series in powers of the coupling,  exploits in an optimal way the information about the high  orders of perturbation theory.  As discussed in detail in \cite{CaFi2}, the method separates the problem of  convergence of the perturbation series from that of its ambiguity, solved by choosing a prescription, which is included in the definition the expansion functions (we adopt here the Principal Value).

In the present paper we worked out in detail the CI and FO versions of the new perturbation expansion for the Adler function in massless QCD. Also, a novelty is the incorporation of the singular behaviour of the Borel transform near the first branch points by factors in the new variable $w$, as shown in (\ref{Bwgamma}) and (\ref{BwgammaFO}). 

In Section \ref{sec:BJ} we illustrated the power of the new perturbation theory using as an example the model of Beneke and Jamin \cite{BeJa}. As expected, the new CIPT proves to be superior, approaching the exact result to a very good accuracy  when the perturbative order increases. The limitations of FOPT due to large imaginary logarithms along the integration contour in the complex plane are clearly illustrated. 

 From the predictions (\ref{newCI}) and  (\ref{newFO}), adding an uncertainty of of 0.003 due to the power corrections \cite{BeJa},   we obtain:
\bea\label{new}
&&\alpha_s(m_\tau^2)=  0.3198\, ^{+ 0.0113}_{-0.0094}\,, \quad   {\rm new ~CIPT },\nonumber\\
&& \alpha_s(m_\tau^2)=  0.3113\,^{+0.0114}_{- 0.0050}\,, \quad {\rm new~ FOPT }.
\eea
As discussed in Section \ref{sec:alphas}, the dominant contribution to the error is due  to the uncertainty of the last perturbative term  in the case of CIPT, and  to the ambiguity of the renormalization scale in the case of FOPT.

It is remarkable that the difference between the central values in (\ref{new})  is  only $0.009$, while for  the standard expansion the difference is $0.024$. The new expansions  remove thus the most intriguing theoretical discrepancy  in the determination of  $\alpha_s$ from $\tau$ decays.  We note that both values  in (\ref{new}) are closer to the standard FOPT than to the standard CIPT. So, our analysis indirectly confirms the criticism  of the standard CIPT made in \cite{BeJa}: although the renormalisation group summation is more accurate, the low order perturbative terms are not able to describe the high order features of the series. The new CIPT brings an improvement precisely at this point. 

According to the last remark made in Section \ref{sec:disc}, for the model considered in \cite{BeJa} the new CIPT and FOPT with $N=5$ terms give comparable predictions, which approximate the true result  from opposite sides. If this model describes correctly the physical situation, then the true result is expected to
 be between  the two predictions  in (\ref{new}). With this assumption, one may take the weighted  average of these two values, which leads to
\be\label{average}
\alpha_s(m_\tau^2)=  0.315\,\, ^{+0.008}_{-0.004}\,, \quad {\rm average}. 
\ee
However, in order to avoid any bias related to a specific model, we take as best result the value given in (\ref{new}) by the new CIPT:
\be\label{final}
\alpha_s(m_\tau^2)=  0.320\, ^{+ 0.011}_{-0.009}\,.
\ee
We recall that this prediction is based on the new contour-improved  expansion defined in Eq. (\ref{DnewCI}). This expansion reproduces the known perturbative coefficients $K_n$ order by order, includes the information about the dominant singularities of the Adler function in the Borel plane, and is based on an optimal expansion of the Borel transform, which converges in the whole complex  $u$-plane up to the cuts along  the real axis. 

  The result  (\ref{final}) coincides practically with that obtained in \cite{BeJa} using the standard FOPT. However, our prediction  has  a more solid theoretical basis, being free of fortuitous compensations of terms related to large imaginary logarithms, like in FOPT. Also, it is based on a systematic perturbation theory,  and its uncertainty is related mainly to the error of the last perturbative term. So, the accuracy  of the prediction is expected to increase when more perturbative terms for the Adler function in QCD will be available.

\vskip1cm
{\bf Acknowledgements:} We are grateful to M. Jamin for providing us the RG-dependent coefficients of the expansion (\ref{astaylor})  and for very useful discussions. One of us (I.C.)  thanks Prof. J.  Ch\'yla  for hospitality at the Institute of Physics of the Czech Academy in Prague.  This work was supported by the  Romanian PN-II Programs  Capacit\u a\c ti of ANCS (Contract No. 15 EU/2009) and Idei of CNCSIS (Contract No. 464/2009),  and by the Projects No. LA08015 of the  Ministry of Education and AV0-Z10100502 of the Academy of Sciences of the Czech Republic.

\end{document}